\documentclass[12pt]{article}
\usepackage{t1enc}
\usepackage{amsmath}
\usepackage{amsthm}
\usepackage{latexsym}
\usepackage{natbib}
\usepackage{epsfig}
\usepackage{textcomp}
\usepackage{alltt}
\usepackage{graphicx}
\usepackage{rotating}
\usepackage{dcolumn}
\usepackage{enumitem}
\usepackage{amsfonts}
\usepackage{algorithm}
\usepackage{amssymb}
\usepackage{bbm, dsfont}
\usepackage[svgnames]{xcolor}
\usepackage{mathtools}
\usepackage{listings}

\usepackage{pdflscape}
\usepackage{afterpage}
\usepackage{capt-of}% or use the larger `caption` package

\usepackage{tikz}
\tikzstyle{ov}=[shape=rectangle,
                draw=black!50,
                thick,
                minimum width=0.7cm,
                minimum height=0.7cm]

\tikzstyle{av}=[shape=rectangle,
                draw=black!50,
                fill=black!10,
                thick,
                minimum width=0.7cm,
                minimum height=0.7cm]

\tikzstyle{lv}=[shape=circle,draw=black!50,thick]

\usetikzlibrary{shapes,calendar,matrix,backgrounds,folding,snakes}
\allowdisplaybreaks

\DeclareMathOperator*{\argmin}{arg\,min}

\begin{document}
\newcommand{\cip}{\perp\!\!\!\!\perp}
\newcommand{\nothere}[1]{}
\newcommand{\noi}{\noindent}
\newcommand{\mbf}[1]{\mbox{\boldmath $#1$}}
\newcommand{\cond}{\, |\,}
\newcommand{\hO}[2]{{\cal O}_{#1}^{#2}}
\newcommand{\hF}[2]{{\cal F}_{#1}^{#2}}
\newcommand{\tl}[1]{\tilde{\lambda}_{#1}^T}
\newcommand{\la}[2]{\lambda_{#1}^T(Z^{#2})}
\newcommand{\I}[1]{1_{(#1)}}
\newcommand{\cd}{\mbox{$\stackrel{\mbox{\tiny{\cal D}}}{\rightarrow}$}}
\newcommand{\cp}{\mbox{$\stackrel{\mbox{\tiny{p}}}{\rightarrow}$}}
\newcommand{\cas}{\mbox{$\stackrel{\mbox{\tiny{a.s.}}}{\rightarrow}$}}
\newcommand{\ld}{\mbox{$\; \stackrel{\mbox{\tiny{def}}}{=3D} \; $}}
\newcommand{\nk}{\mbox{$n \rightarrow \infty$}}
\newcommand{\con}{\mbox{$\rightarrow $}}
\newcommand{\dprime}{\mbox{$\prime \vspace{-1 mm} \prime$}}
\newcommand{\Borel}{\mbox{${\cal B}$}}
\newcommand{\bevis}{\mbox{$\underline{\em{Proof}}$}}
\newcommand{\Rd}[1]{\mbox{${\Re^{#1}}$}}
\newcommand{\il}[1]{{\int_{0}^{#1}}}
\newcommand{\pl}[1]{\mbox{\bf {\LARGE #1}}}
\newcommand{\expit}{\text{expit}}
\newcommand{\indep}{\rotatebox[origin=c]{90}{$\models$}}
\newcommand{\blind}{1}
\newcommand{\pr}{\text{pr}}
\newcommand{\var}{\text{var}}
\newcommand{\cov}{\text{cov}}
\newcommand{\Bin}{\text{Bin}}
\newcommand{\Exp}{\text{Exp}}
\newcommand{\unif}{\text{unif}}
\newcommand{\logit}{\text{logit}}
\newcommand{\sign}{\text{sign}}
\newcommand{\support}{\text{support}}
\newcommand\norm[1]{\left\lVert#1\right\rVert}

\newtheorem{theorem}{Theorem}
\newtheorem{lemma}{Lemma}
\newtheorem{prop}{Proposition}
\newtheorem{assumption}{Assumption}
\newtheorem{definition}{Definition}
\newtheorem*{remark}{Remark}
\newtheorem{corollary}{Corollary}
\newtheorem{example}{Example}

\parindent12pt

\begin{center}{\Large{On Estimation and Cross--validation of Dynamic Treatment Regimes with Competing Risks}}
 \end{center}
 
 { 
\begin{center}
Pawe\l{} Morzywo\l{}ek$^{1}$, Johan Steen$^{2,3,4}$, Wim Van Biesen$^{2,3}$, \\
Johan Decruyenaere$^{2,4}$, Stijn Vansteelandt$^{1,5}$\\
\bigskip
\scriptsize{$^1$Department of Applied Mathematics, Computer Science and Statistics, Ghent University} \\
\scriptsize{$^2$Department of Internal Medicine and Pediatrics, Ghent University} \\
\scriptsize{$^3$Renal Division, Ghent University Hospital} \\
\scriptsize{$^4$Department of Intensive Care Medicine, Ghent University Hospital} \\
\scriptsize{$^5$Department of Medical Statistics, London School of Hygiene and Tropical Medicine}
\end{center}
}
\smallskip
\begin{center}
\today
\end{center}
\smallskip

\setlength{\parindent}{0.3in} \setlength{\baselineskip}{24pt}
\begin{abstract}
The optimal moment to start renal replacement therapy in a patient with acute kidney injury (AKI) remains a challenging problem in intensive care nephrology. Multiple randomised controlled trials have tried to answer this question, but these contrast only a limited number of treatment initiation strategies. In view of this, we use routinely collected observational data from the Ghent University Hospital intensive care units (ICUs) to investigate different prespecified timing strategies for renal replacement therapy initiation based on time--updated levels of serum potassium, pH and fluid balance in critically ill patients with AKI with the aim to minimize $30$--day ICU mortality. For this purpose, we apply statistical techniques for evaluating the impact of specific dynamic treatment regimes in the presence of ICU discharge as a competing event. We discuss two approaches, a nonparametric one -- using an inverse probability weighted Aalen--Johansen estimator -- and a semiparametric one -- using dynamic--regime marginal structural models. Furthermore, we suggest an easy to implement cross--validation technique to assess the out--of--sample performance of the optimal dynamic treatment regime. Our work illustrates the potential of data--driven medical decision support based on routinely collected observational data.
\end{abstract}
\noi
{\it Dynamic Treatment Regimes, Cross--validation, Competing Events, Treatment--Confounder Feedback, Marginal Structural Models, Aalen--Johansen Estimator, Precision Medicine, Acute Kidney Injury, Renal Replacement Therapy}

\section{Introduction}

Digitalization of patient records and increasing computational power have led to a paradigm shift in the field of medical decision--making from one--size--fits--all interventions to more sophisticated and data--driven intervention strategies optimised for particular subpopulations or individuals. This transition towards precision medicine \citep{Ashley2015, Kosorok2019} has  fuelled expectations to successfully tackle some challenging yet unresolved medical problems. One such unresolved question in intensive care medicine is when to initiate renal replacement therapy (RRT, dialysis) in patients with acute kidney injury (AKI). While RRT can avert some of the life--threatening consequences of kidney failure, it is also an invasive treatment and might put patients at increased risk of adverse events. Despite multiple observational studies and clinical trials comparing immediate treatment initiation with strategies that delay treatment initiation until one of the so--called "hard criteria" (sometimes also called "absolute indications") has been met, the question of optimal timing of RRT initiation remains subject of discussion \citep{Gaudry2016, Zarbock2016, Vanmassenhove2017}. 
\\
\indent
The need for individualized decision support has led causal inference researchers to investigate so called \emph{dynamic treatment regimes} (DTR) \citep{Chakraborty2013, Tsiatis2020}. A \emph{dynamic treatment regime} is a set of sequential rules for treatment decisions based on patients' time--varying covariates, e.g. "start RRT in AKI--diagnosed patients if pH fell below $7.1$ in the last $24$h". In contrast, a static treatment regime does not take into account post--baseline information about the patient, e.g. "always initiate RRT within $24$h from the AKI diagnosis" or "never initiate RRT in AKI--diagnosed patients". Some of the early work in the field of dynamic treatment regimes introduced regret--based learning (sometimes referred to as \emph{A--learning}) \citep{Murphy2003} and g--estimation \citep{Robins2004}. These methods allow to find flexible optimal decision rules from a rich set of potential treatment strategies. A parallel line of research aimed to constrain the search for optimal treatment rules to a more limited set of simple and well--interpretable treatment strategies. Some of the most prominent methods applied in this context are based on \emph{marginal structural models} \citep{Murphy2001, VanderLaan2007, Orellana2010}. 
\\
\indent
The main motivation for our work is to investigate different prespecified treatment strategies for initiation of renal replacement therapy to minimize $30$--day ICU mortality in critically ill patients with stage $2$ AKI. There are three main complications in the analysis. The first difficulty is that the main event of interest -- \emph{ICU death within $30$ days from AKI diagnosis} -- is subject to a competing event, namely alive discharge from the ICU (henceforth referred to as \emph{ICU discharge}). The second challenge relates to the time--varying nature of treatment decisions and of confounding. In practice, past information on patients' evolving condition and kidney function is used to make a decision on whether to initiate treatment, which in turn influences future kidney function. This induces treatment--confounder feedback that cannot be controlled for via traditional methods to deal with confounding, like stratification, matching or regression adjustment \citep{Hernan2020}. The third challenge is reliable performance assessment and comparison of considered treatment regimes. The commonly used approach is to evaluate all regimes on the full data, choose the one with the best performance (in our example the one with the lowest $30$--day ICU mortality) and report its performance on the same data set.  Such in--sample performance evaluation may lead to an overly optimistic assessment. 
\\
\indent
This paper provides guidance on how competing events and treatment--confounder feedback can adequately be taken into account when comparing different dynamic treatment strategies. In particular, we focus on a set of treatment strategies, which involve mild deviations from current clinical practice, and evaluate their comparative effectiveness with the aim to identify an optimal strategy that improves outcomes compared to current clinical practice. We address the question of optimal timing of RRT initiation using a semiparametric estimation approach based on \emph{dynamic--regime marginal structural models} (DYR--MSM) \citep{Murphy2001, VanderLaan2007, Orellana2010} in the context of competing risks survival analysis and compare it with a weighted version of the nonparametric Aalen--Johansen estimator \citep{Aalen1978} applied to dynamic treatment regimes. Furthermore, we apply cross--validation for out--of--sample performance assessment of the estimated optimal dynamic treatment regime. Cross--validation methodology for the performance evaluation of statistical procedures is commonly applied in the statistical prediction literature with the aim to avoid overoptimism \citep{Stone1974,  Harrell1996, Steyerberg2001, Dudoit2005, Hastie2009, Smith2014, VanderLaan2014}. 
\\
\indent
The paper is structured as follows. In Section \ref{sec2}, we describe the problem setting and introduce the key concepts related to dynamic treatment regimes and competing risks survival analysis. In Section \ref{sec3}, we describe the use of cross--validation for the out--of--sample performance evaluation of the optimal dynamic treatment regime. In Section \ref{sec4}, we discuss different approaches to estimate cumulative incidence curves under specific dynamic treatment regimes. In Section \ref{sec5}, we present an empirical data application of the discussed methodology for optimizing the time of treatment initiation in the AKI patient population. In Section \ref{sec6}, we provide a discussion of the results. 

\section{Problem Setting and Notation}\label{sec2}

We consider $K \geq 1$ time points at which treatment decisions need to be made. In our setting $K = 30$, since we are interested in $30$--day ICU mortality and treatment initiation is considered daily. Furthermore, throughout the main text we assume no censoring due to administrative end of the study or loss to follow--up. However we describe the extension of the methods incorporating censoring due to administrative end of the study or loss to follow--up in Appendix \ref{app0}. In the paragraphs that follow we denote random variables with capital letters and their potential realizations by small letters. In particular, let $A_k$, $k=0,\dots, K-1$, denote a treatment option at decision time $k$, which takes value $a_k \in \{0,1\}$, with $1$ indicating that RRT has been initiated and $0$ otherwise. Let $X_0$ be the information available about the individual at the time of stage $2$ AKI diagnosis (which is considered as start of follow--up, or time $0$), including baseline characteristics. Similarly, let $X_k$, $k=1,\dots,K$ denote patient information that becomes available between decision times $k-1$ and $k$, e.g. the lowest level of pH in the last 24h, etc. Let $Y_k$, $k=0,\dots,K$ indicate whether or not the \emph{event of interest (ICU death)} has been observed by time $k$ (with $Y_k = 1$ if it has or $Y_k = 0$ otherwise). Similarly, let $Z_k$, $k=0,\dots,K$ indicate whether or not the \emph{competing event (ICU discharge)} has been observed by time $k$ (with $Z_k = 1$ if it has or $Z_k = 0$ otherwise). Note that only one of two events can be observed for a single patient. We assume the following order of the variables within the $k^{th}$ time interval $\left( Z_k, Y_k, X_k, A_k \right)$. The ordering needs to be aligned with our understanding of the data--generating mechanism and consistently applied across time intervals. Furthermore, define $O_k \equiv \left( Z_k, Y_k, X_k \right)$ and for a time--varying variable $H_k$, $k = 0, 1, \dots, K$, let $\overline{H}_k \equiv \left( H_0, \dots, H_k \right)$ be the history until and including time $k$.  For all patients in the study population $A_k = 0$ for $k < 0$, $Y_k = 0$ for $k \leq 0$ and $Z_k = 0$ for $k \leq 0$. 
\\
\indent
A \emph{decision rule} at time $k$ is a function that maps the individual's history to a treatment option $d_k \colon \left( \overline{o}_k, \overline{a}_{k-1} \right) \rightarrow a_k$, e.g. "start RRT in AKI--diagnosed patients if pH fell below $7.1$ in the last 24h". Furthermore, we define $\overline{d}_k \left( \overline{o}_k, \overline{a}_{k-1} \right) \equiv \left( d_0 \left( o_0, a_{-1} \right), \dots, d_k \left( \overline{o}_k, \overline{a}_{k-1} \right)\right)$. A \emph{dynamic treatment regime} is a collection of decision rules, one decision rule corresponding with each of the $K$ decision points, i.e. $d \equiv \overline{d}_{K-1}$. 
\\
\indent
For the purpose of defining and identifying the optimal treatment regime we use the potential outcomes framework from the causal inference literature \citep{Neyman1923, Rubin1978, Robins1986}. Let $Y^d_k$, $k=1,\dots,K$ be the \emph{potential outcome for the event of interest (ICU death) at time k} that would be observed if all individuals had followed the dynamic treatment regime $d$ until time $k-1$.  Analogously, let $Z^d_k$, $k=1,\dots,K$ be the \emph{potential outcome for the competing event (ICU discharge) at time k} that would be observed if all individuals had followed the dynamic treatment regime $d$ until time $k-1$. Note that the considered framework merely infers aspects of the distribution under a single, fixed treatment regime, hence alternative frameworks (e.g. decision--theoretic) to infer the effect of dynamic treatment strategies could be used \citep{Dawid2010}.  
\\
\indent
Let $\mathcal{D} = \left\lbrace d^{(1)}, \dots, d^{(m)} \right\rbrace$ denote a set of prespecified dynamic treatment regimes. We aim to identify the \emph{optimal regime} $d_{opt, v} \in \mathcal{D}$ for a subpopulation with characteristics $V=v$ for $V \subset X_0$, a possibly empty subset of the baseline covariates. To compare different treatment regimes in the search for the optimal one, we define the \emph{value} of a regime $d$ (for the subpopulation with characteristics $V=v$) to be $30$--day ICU mortality $\mathcal{V} \left(d; V=v \right) \equiv \mathbb{P} \left( Y^d_K = 1 \vert V = v \right)$. Furthermore, we define the \emph{optimal regime} as the regime that minimizes the value, i.e. $d_{opt, v} \equiv \argmin_{d \in \mathcal{D}} \mathcal{V} \left(d; V = v \right)$. 
\\
\indent
To estimate $\mathbb{P} \left( Y_{k}^d = 1 \vert V = v \right)$, for $k = 1, \dots, K$, we define the \emph{counterfactual event--specific hazard}, as the event--specific hazard of an event at time $k$ under treatment regime $d$ among patients with characteristics $V = v$ who have not previously experienced the event of interest nor the competing event. Using the definitions introduced above the counterfactual discrete--time event--specific hazard for the event of interest under the treatment regime $d$ at time $k$ can be written as 
\begin{equation}
\label{eqn1}
h_{k}^{(1)} \left(d, V \right) \equiv \mathbb{P} \left( Y_{k}^d = 1 \vert Y_{k-1}^d = 0, Z_{k}^d = 0, V \right).
\end{equation}
Similarly, the counterfactual event--specific hazard for the competing event under the treatment regime $d$ at time $k$ is defined as
\begin{equation}
\label{eqn2}
h_{k}^{(2)} \left(d, V \right) \equiv \mathbb{P} \left( Z_{k}^d = 1 \vert Z_{k-1}^d = 0, Y_{k-1}^d = 0, V \right).
\end{equation}
Even in the absence of censoring due to loss to follow--up, we advocate a hazard--based approach for estimating $\mathbb{P} \left( Y^d_k = 1 \vert V = v \right)$, for $k = 1, \dots, K$, instead of direct estimation of the risk of the binary outcome $Y^d_k$. This choice is further motivated in Section \ref{subsection:label3}. 
\\
\indent
It is important to notice that these hazards, while defined in terms of counterfactuals, do not carry a causal interpretation due to a built--in selection bias when comparing two populations defined post--baseline. Indeed, imagine we would like to compare the effect of treatment regime $d^{(1)}$ relative to $d^{(2)}$ using time--varying hazard (\ref{eqn1}). Furthermore, assume that the treatment strategy $d^{(1)}$ is detrimental to a frail subpopulation of patients. For the sake of the argument we assume that the patient population is randomly assigned to one of the treatment regimes at baseline, however the argument generalizes to observational studies. Due to randomization the proportion of frail patients in both groups will be similar at baseline allowing fair comparison of hazards in both groups at time 0. However, as time progresses the proportion of frail patients in the cohort assigned to $d^{(1)}$ will decrease due to the detrimental effect of this treatment strategy. This means that at later time points the hazard rate in the group assigned to $d^{(1)}$ will be computed over a healthier population than in those assigned to $d^{(2)}$, hence leading to a potentially misleading comparison \citep{Hernan2010, Martinussen2018, Young2020}. 
\\
\indent
To avoid such selection bias, we instead compare cumulative incidence functions. These can be expressed as functionals of the counterfactual event--specific hazards by the following formula for $k=1,\dots,K$ \citep{Young2020}
\begin{equation}
\label{eqn3}
\mathbb{P} \left( Y_k^d = 1 \vert V \right) = \sum_{l=1}^{k} h_l^{(1)} \left(d, V \right) \left\lbrace 1 - h_l^{(2)} \left(d, V \right) \right\rbrace \prod_{j=1}^{l-1} \left\lbrace 1-h_j^{(1)} \left(d, V \right) \right\rbrace \left\lbrace 1-h_j^{(2)} \left(d, V \right) \right\rbrace.
\end{equation}
This formula expresses that the probability of observing the event of interest by day $k$ equals the sum of probabilities to observe the event of interest by any particular day, but not the competing event on that day (following the assumed order of events within a time interval) and also no event on any of the previous days. 

\section{Cross--validation for Dynamic Treatment Regimes}\label{sec3}

The purpose of using cross--validation in our analysis is to assess the best performing dynamic treatment regime on a separate set than the one used to select it. This is encouraged by the TRIPOD (Transparent reporting of a multivariable prediction model for individual prognosis or diagnosis) \citep{Collins2015} guidelines for internal model validation. In particular, we wish to evaluate what would be the performance of the optimal DTR from a predefined set of potential treatment regimes $\mathcal{D} = \left\lbrace d^{(1)}, \dots, d^{(m)} \right\rbrace$ when deployed in a clinical environment. The commonly used approach is to evaluate all regimes in $\mathcal{D}$ on the full data, choose the one with the best performance (in our example the one with the lowest $30$--day ICU mortality) and report its performance on the same data set.  Such in--sample performance evaluation may lead to an overly optimistic assessment. For instance, imagine we would like to compare two treatment regimes $d^{(1)}$ and $d^{(2)}$. Furthermore, assume both regimes have exactly the same performance. However, once we evaluate and compare them on our data, then merely due to sampling variability, one may seem to perform better than the other. Claiming the optimality of this regime and reporting its in--sample performance would then leave a false optimistic impression of the superiority of one regime.
\\
\indent
To perform an out--of--sample evaluation, one could split the data into a training set and a validation set and then evaluate all DTRs on the training set to pick the one with the lowest ICU mortality.  Once this has been done, one could report the out--of sample performance of this best performing DTR as calculated on the test set. This will result in a less optimistic evaluation than that based on the in--sample performance of the DTR on the training set. It reflects the expected performance of the data--driven procedure to select an optimal DTR when deployed in real--time in a clinical environment.  We apply an out--of--sample performance assessment based on cross--validation. This extends the idea of performance evaluation done on the test set by repeating this procedure using different data splits. 
\\
\indent
We perform $J$--fold cross--validation by splitting the available data set into $J$ mutually exclusive parts of equal size. We combine $J-1$ folds into a training set and use the remaining $J^{th}$ fold as a test set. We permute the role of different folds, so that each fold becomes the test set exactly once. For $j \in \left\lbrace 1, \dots, J \right\rbrace$, let $\hat{\mathbb{P}}_{j}^{train} \left( Y_{K}^d = 1 \vert V = v \right)$ be $30$--day ICU mortality for the subpopulation with characteristics $V=v$ under the dynamic treatment regime $d$ estimated on the $j^{th}$ training set and $\hat{\mathbb{P}}_{j}^{test} \left( Y_{K}^d = 1 \vert  V = v \right)$ be $30$--day ICU mortality (for the same subpopulation) under the dynamic treatment regime $d$ estimated on the $j^{th}$ test set. We estimate the cumulative incidence curve up to time $K$ on the training set for each considered dynamic treatment regime using the methods described in the following sections. For each $j \in \left\lbrace 1, \dots, J \right\rbrace$, we choose the best treatment regime $\hat{d}_{opt, v}^{(j)}$ in the $j^{th}$ training set to be the one attaining the lowest $30$--day ICU mortality on the $j^{th}$ training set
\begin{equation*}
\hat{d}_{opt, v}^{(j)} \equiv \argmin_{d \in \mathcal{D}} \hat{\mathbb{P}}_{j}^{train} \left( Y_{K}^d = 1 \vert  V = v \right).
\end{equation*}
This leads to $J$ estimated optimal dynamic treatment regimes $\hat{d}_{opt, v}^{(1)}, \dots, \hat{d}_{opt, v}^{(J)}$, together with the estimated $30$--day ICU mortality for the optimal rule on the $j^{th}$ test set
\begin{equation*}
\hat{\mathbb{P}}_{j}^{test} \left( Y_{K}^{\hat{d}_{opt, v}^{(j)}} = 1 \vert  V = v \right)
\end{equation*} 
for $j = 1, \dots, J$. We then compute the cross--validated counterfactual $30$--day ICU mortality under the optimal regime given by
\begin{equation}
\label{eqn4}
\frac{1}{J} \sum_{j=1}^J \hat{\mathbb{P}}_{j}^{test} \left( Y_{K}^{\hat{d}_{opt, v}^{(j)}} = 1 \vert  V = v \right).
\end{equation}
It is important to note that the optimal $\hat{d}_{opt, v}^{(j)}$ corresponding to different folds $j = 1, \dots, J$ might not always be the same regime. In particular, (\ref{eqn4}) may not correspond to the $30$--day ICU mortality of a single DTR,  but it can be interpreted as $30$--day ICU mortality that would have been obtained by deploying the procedure of choosing the optimal treatment regime out of set $\mathcal{D}$ of possible rules on new data. The benefit of the cross--validated counterfactual $30$--day ICU mortality given by (\ref{eqn4}) is twofold. By comparing it with the cumulative incidence under current standard of care we assess whether deploying the optimal regime is expected to lead to a decrease in the ICU mortality. Furthermore, by comparing the cross--validated mortality with the $30$--day ICU mortality under the optimal treatment regime evaluated on the whole data set, we obtain an assessment of the optimism due to selection and evaluation of the optimal treatment regime on the same data set.  Note that this assessment may be somewhat conservative (i.e. overestimated) due to the reduction in size of the training set.
\\
\indent
Cross--validation has also been suggested \citep{Chakraborty2014} or considered \citep{Luedtke2018} by others. For instance, \cite{Luedtke2018} rely on a cross--validation procedure like the above to obtain data--adaptive estimates of the mean outcome under an estimated optimal regime, along with confidence intervals.  

\section{Estimation of the Cumulative Incidence Curve}\label{sec4}

We discuss two approaches to estimate the cumulative incidence curve given by (\ref{eqn3}) -- a nonparametric one, leading to the inverse probability of censoring weighted (IPC weighted) Aalen--Johansen estimator of the cumulative incidence \citep{Aalen1978, Young2020} and a semiparametric one, using dynamic--regime marginal structural models (DYR--MSM) \citep{Murphy2001, VanderLaan2007, Orellana2010}. 
\\
\indent
In our motivating example, we are interested in investigating the optimal threshold $x$ for a decision rule "start RRT in AKI--diagnosed patients if pH fell below $x$ in the last 24h". The IPC weighted Aalen--Johansen estimator targets the cumulative incidence curve for a single treatment regime, independently of the other regimes. The advantage of a dynamic--regime marginal structural model is that it permits to postulate a joint model for different treatment regimes corresponding to different threshold values.  This may be appealing when a data set carries little information about certain dynamic treatment regimes and smoothing across similar regimes is warranted. 

\subsection{Dynamic Treatment Regime Compatibility}\label{subsection:label2}

Given our aim to identify the best performing treatment strategy $d_{opt,v}$ out of the set of prespecified treatment strategies $\mathcal{D} = \left\lbrace d^{(1)}, \dots, d^{(m)} \right\rbrace$, the favoured way to compare different DTRs would be to perform an $m$--arm randomized trial in which patients are randomly assigned to one of the considered treatment regimes in $\mathcal{D}$. Unfortunately, once the number of treatment regimes grows, such a randomized trial becomes infeasible. Alternatively, we can evaluate each of the treatment strategies in $\mathcal{D}$ using observational data \citep{Cain2010, Hernan2016} and compare their performance to identify the best one. The difficulty that we are facing in this case is the fact that we are interested in evaluating retrospectively what would be the effect of applying a hypothetical treatment strategy $d \in \mathcal{D}$ in the study population, whereas the patients included in the observational data set, had been treated based on the treatment strategy followed by their clinicians, which might have been very different from the dynamic treatment strategy $d$ we wish to evaluate. Therefore, to evaluate performance of the hypothetical treatment strategy $d$ we need to identify patients in the data set whose observed treatment history coincides with the treatment trajectory that would be observed under the hypothetical treatment regime $d$ we wish to evaluate, because only these patients can provide information on the effect of hypothetical DTR of interest. We will refer to these patients as compatible with the dynamic treatment regime $d$ that we consider. Note that all patients are compatible at time $0$ with any of the considered regimes $d$. Now we follow each individual across time and we artificially censor him/her as soon as the observed treatment trajectory deviates from the one that would be observed under the considered DTR $d$. In other words, we artificially censor a patient, if and as soon as that patient stops being compatible with the considered DTR.  We define the indicator of being compatible with the treatment regime $d$ at time $k$ as 
\begin{equation*}
C_{d,k} \equiv \mathbbm{1} \left\lbrace \overline{A}_k = \overline{d}_k \left( \overline{O}_k, \overline{A}_{k-1} \right) \right\rbrace
\end{equation*}
where
\begin{equation*}
\mathbbm{1} \left( X = x \right) = 
\begin{cases}
 1,   & X=x\\
 0,  & \text{otherwise}
\end{cases}
\end{equation*}
is an indicator function. Therefore, a patient is compatible with the considered treatment regime $d$ at time $k$, if her/his treatment history coincides with the treatment options that would have been assigned by $d$ for all the past decision points. The idea of estimating the effect of a treatment regime through censoring individuals the first time their treatment differs from the treatment assigned by the considered DTR has been commonly used in the dynamic treatment regimes literature \citep{Robins1993, Murphy2001, VanderLaan2007, Orellana2010, Cain2010, Chakraborty2013,Tsiatis2020}.
\\
\indent
In order to find the optimal treatment regime we need to evaluate each of the DTRs in $\mathcal{D}$ and compare them to identify the one with the lowest resulting mortality.  For this purpose we create an extended data set with one copy of each subject in the original observational data set for each considered treatment regime. For each regime we follow the patient population over time and we artificially censor individuals as soon as they are no longer compatible with that given regime. The probability of remaining compatible with the dynamic treatment regime $d$ at time $k$ (conditional on being compatible up to time $k-1$) is given by 
\begin{align}
\label{eqn5}
\nonumber
&\pi_{d,k} \equiv \mathbb{P} \left\lbrace C_{d,k} = 1 \vert \overline{O}_k, C_{d,k-1} = 1 \right\rbrace \\
&=
\begin{cases}
\mathbb{P} \left\lbrace C_{d,k} = 1 \vert \overline{X}_k,  \overline{Y}_k = \overline{0}, \overline{Z}_k = \overline{0}, C_{d,k-1} = 1 \right\rbrace, \text{if } \overline{Y}_k = \overline{0} \text{ and } \overline{Z}_k = \overline{0} \\
1, \text{ otherwise.}
\end{cases}
\end{align}
Therefore, the probability of being artificially censored with respect to the treatment regime $d$ at time $k$ (conditional on being compatible up to time $k-1$) is given by $1 - \pi_{d,k}$.  Note that if one of the competing events occurs at time $k$ (i.e. $Y_k=1$ or $Z_k=1$) while an individual is compatible with the dynamic treatment regime $d$ (i.e. $C_{d,k-1} = 1$), then the individual automatically remains compatible for all future time points. 
\\
\indent
As an example consider a data set containing information on two patients observed over 5 days. Suppose we want to compare two dynamic treatment regimes "start RRT in AKI--diagnosed patients if pH fell below $7.1$ in the last $24$h" and "start RRT in AKI--diagnosed patients if pH fell below $7.2$ in the last $24$h". Therefore we make two copies of the original data set -- one for each treatment regime that we consider -- and stack them on top of each other to obtain an extended data set (see Table \ref{table1}). The compatibility indicator in Table \ref{table1} corresponds to the compatibility status of the patient at the end of the day. Given that a patient is compatible with a considered regime $d$ at time $k$, we can determine the compatibility indicator on the following day by checking whether the treatment decision based on the dynamic treatment regime coincides with the observed treatment. For the DTR "start RRT in AKI--diagnosed patients if pH fell below $x$ in the last $24$h" a given patient would remain compatible if the minimal level of pH in the last $24$h fell below the threshold $x$ and the treatment has been initiated or if the minimal level of pH did not fall below the threshold $x$ in the last $24$h and treatment has not been started. Once the patient ceases to be compatible with a particular treatment regime at some time point, then he/she will be incompatible with this regime for all future time points. Furthermore, the patients that are compatible with a particular DTR at the time of experiencing one of the events -- ICU death or ICU discharge -- remain compatible with the considered dynamic treatment regime until the end of the study.

\begin{landscape}
\begin{table}[h!]
\begin{center}
\begin{tabular}{c|c|c|c|c|c|c|c|c|c} % <-- Alignments: 1st column left, 2nd middle and 3rd right, with vertical lines in between

\textbf{ID} & \multicolumn{1}{|p{1.3cm}|}{\centering \textbf{Time} \\ \textbf{(days)}} & \textbf{DTR} & \multicolumn{1}{|p{2.1cm}|}{\centering \textbf{Observed} \\ \textbf{Treatment}} & \multicolumn{1}{|p{2cm}|}{\centering \textbf{Min. pH} \\ \textbf{(last $24$h)}} & \multicolumn{1}{|p{2.8cm}|}{\centering \textbf{Compatibility} \\ \textbf{Indicator}} & \textbf{PS} & \textbf{IPCW} & \multicolumn{1}{|p{1.4cm}|}{\centering \textbf{ICU} \\ \textbf{Death}}  & \multicolumn{1}{|p{2.1cm}}{\centering \textbf{ICU} \\ \textbf{Discharge}} \\
      \hline
     
      \hline
1 & 1 & 7.1 & 0 & 7.29 & 1 & 0.01 & 1.01 & 0 & 0 \\
1 & 2 & 7.1 & 0 & 7.24 & 1 & 0.05 & 1.06 & 0 & 0 \\
1 & 3 & 7.1 & 1 & 7.08 & 1 & 0.4 & 2.67 & 0 & 0 \\
1 & 4 & 7.1 & 1 & 7.29 & 1 & 1.0 & 2.67 & 0 & 0 \\
1 & 5 & 7.1 & 1 & 7.29 & 1 & 1.0 & 2.67 & 0 & 1 \\
      \hline
2 & 1 & 7.1 & 0 & 7.3 & 1 & 0.13 & 1.15 & 0 & 0 \\
2 & 2 & 7.1 & 0 & 7.29 & 1 & 0.11 & 1.3 & 0 & 0 \\
2 & 3 & 7.1 & 1 & 7.19 & 0 & 0.48 & 0.0 & 0 & 0 \\
2 & 4 & 7.1 & 1 & 7.32 & 0 & 1.0 & 0.0 & 1 & 0 \\
2 & 5 & 7.1 & 1 & 7.32 & 0 & 1.0 & 0.0 & 1 & 0 \\
      \hline
      \hline
1 & 1 & 7.2 & 0 & 7.29 & 1 & 0.01 & 1.01 & 0 & 0 \\
1 & 2 & 7.2 & 0 & 7.24 & 1 & 0.05 & 1.06 & 0 & 0 \\
1 & 3 & 7.2 & 1 & 7.08 & 1 & 0.4 & 2.67 & 0 & 0 \\
1 & 4 & 7.2 & 1 & 7.29 & 1 & 1.0 & 2.67 & 0 & 0 \\
1 & 5 & 7.2 & 1 & 7.29 & 1 & 1.0 & 2.67 & 0 & 1 \\
      \hline
2 & 1 & 7.2 & 0 & 7.3 & 1 & 0.13 & 1.15 & 0 & 0 \\
2 & 2 & 7.2 & 0 & 7.29 & 1 & 0.11 & 1.3 & 0 & 0 \\
2 & 3 & 7.2 & 1 & 7.19 & 1 & 0.48 & 2.9 & 0 & 0 \\
2 & 4 & 7.2 & 1 & 7.32 & 1 & 1.0 & 2.9 & 1 & 0 \\
2 & 5 & 7.2 & 1 & 7.32 & 1 & 1.0 & 2.9 & 1 & 0 \\
      \hline
      \hline
\end{tabular}
\caption{Extended data set for comparison of two dynamic treatment regimes "start RRT in AKI--diagnosed patients if pH fell below $7.1$ in the last $24$h" and "start RRT in AKI--diagnosed patients if pH fell below $7.2$ in the last $24$h".  "Min. pH (last $24$h)" is the lowest value of pH measured in the last $24$h for a given patient.  "PS" is propensity score given by (\ref{eqn7}) and "IPCW" is inverse probability of censoring weight given by (\ref{eqn6}). } \label{table1}
\end{center}
\end{table}
\end{landscape}

\subsection{Identification of the Causal Estimands}

As discussed in the previous section, to compare different dynamic treatment regimes $d$ we need to estimate $\mathbb{P} \left( Y_K^d = 1 \vert V = v \right)$. Because this is a counterfactual estimand, it cannot be estimated from the data without invoking common assumptions from the causal inference literature. The consistency assumption is crucial for the application of the potential outcomes framework as it links the observed data to the potential outcomes. It states that for each patient compatible with dynamic treatment regime $d$ up until time $k$ the observed variables at time $k$ correspond to the potential outcome variables under treatment regime $d$. This is expressed for $k = 0, \dots, K-1$ as
\begin{equation*}
\overline{O}_{k+1}^d = \overline{O}_{k+1} \hspace{0.3cm} \text{if} \hspace{0.3cm} C_{d,k} = 1,
\end{equation*}
where $O_{k}^d \equiv \left( Z_{k}^d, Y_{k}^d, X_{k}^d \right)$ and $X_{k}^d$ is the potential outcome for patient information that would become available between decision times $k-1$ and $k$ under regime $d$. Another important assumption is sequential randomization, which can be stated as follows 
\begin{equation*}
\left( \underline{Z}_{k+1}^d,  \underline{Y}_{k+1}^d \right) \cip A_k \vert \overline{O}_k,  \overline{A}_{k-1} = \overline{d}_{k-1} \left( \overline{O}_{k-1}, \overline{A}_{k-2} \right), \text{ for } k = 0, \dots, K-1,
\end{equation*}
where $\underline{Y}_{k}^d \equiv \left( Y_k, \dots, Y_K \right)$ and $\underline{Z}_{k}^d \equiv \left( Z_k, \dots, Y_K \right)$, for $k = 1, \dots, K$, denote the future trajectory of the potential outcomes for the event of interest and competing event, respectively,  if all individuals had followed the dynamic treatment regime $d$. This assumption holds by design if treatment initiation is randomly assigned at each decision point $k$ among patients in whom RRT has not been initiated before time $k$, with randomization possibly depending on the observed past. However, in observational studies this assumption requires careful consultation with subject matter experts as it essentially states that there are no unmeasured variables that influence both the treatment assignment at a given time point and the final outcome, i.e. no unmeasured confounding. Finally, the positivity assumption for the dynamic treatment regimes states that for $k = 0, \dots, K-1$ 
\begin{equation*}
\mathbb{P} \left\lbrace A_k = d_k \left( \overline{O}_k,  \overline{A}_{k-1} \right) \vert \overline{X}_k, \overline{Y}_k = 0, \overline{Z}_k = 0,  C_{d,k-1} = 1 \right\rbrace > 0, \text{ w.p.1,}  
\end{equation*}
i.e. each patient that has remained event--free and compatible with the treatment regime $d$ up until time $k$ has a positive probability of receiving treatment prescribed by $d_k$ at time $k$.  

\subsection{Inverse Probability of Censoring Weighting}

Artificial censoring due to incompatibility with a treatment regime $d$ at time $k$ might select a group of patients that is no longer comparable with the risk set under the observed regime in terms of prognostic factors for the outcome. One approach to remedy selection bias due this artificial censoring is the use of inverse probability of censoring weighting (IPC weighting). IPC weights are given by
\begin{equation}
\label{eqn6}
W_{d,k,i} \equiv \frac{C_{d,k,i}}{ \prod_{l=1}^k \pi_{d,l,i}},
\end{equation}
where $\pi_{d,l,i}$ is the probability that an individual $i$ remains compatible with the DTR $d$ by the time $l$ conditional on the observed history and on having remained compatible up to the previous time point given by (\ref{eqn5}). To obtain the estimates $\hat{\pi}_{d,k,i}$, and hence $\hat{W}_{d,k,i}$, we note first that the upper part of the expression (\ref{eqn5}) can be written as
\begin{align*}
&\mathbb{P} \left\lbrace A_k = d_k \left( \overline{O}_k, \overline{A}_{k-1} \right) \vert \overline{X}_k,  \overline{Y}_k = \overline{0}, \overline{Z}_k = \overline{0}, C_{d,k-1} = 1 \right\rbrace \\
&=\mathbb{P} \left\lbrace A_k = 1 \vert \overline{X}_k,  \overline{Y}_k = \overline{0}, \overline{Z}_k = \overline{0}, C_{d,k-1} = 1 \right\rbrace^{d_k \left( \overline{O}_k, \overline{A}_{k-1} \right)} \\
&\times \left[ 1-\mathbb{P} \left\lbrace A_k = 1 \vert \overline{X}_k,  \overline{Y}_k = \overline{0}, \overline{Z}_k = \overline{0}, C_{d,k-1} = 1 \right\rbrace \right]^{1-d_k \left( \overline{O}_k, \overline{A}_{k-1} \right)}.
\end{align*}
Furthermore, note that for individuals compatible with the DTR $d$ by the time $k$ we have
\begin{align}
\label{eqn7}
\nonumber
&\mathbb{P} \left( A_k = 1 \vert \overline{X}_k,  \overline{Y}_k = \overline{0}, \overline{Z}_k = \overline{0}, C_{d,k-1} = 1 \right) \\
&=
\begin{cases}
	\mathbb{P} \left( A_k = 1 \vert \overline{X}_k,  \overline{Y}_k = \overline{0}, \overline{Z}_k = \overline{0},  \overline{A}_{k-1} = \overline{a}_{k-1} \right),\text{ if } \overline{a}_{k-1} = \overline{0} \\
	1,\text{ otherwise}
\end{cases}
\end{align}
where $\mathbb{P} \left( A_k = 1 \vert \overline{X}_k, \overline{Y}_k = \overline{0}, \overline{Z}_k = \overline{0}, \overline{A}_{k-1} = \overline{a}_{k-1} \right)$ is the time--dependent propensity score, i.e. the probability of receiving treatment at a given time point given the observed event and covariate history. Note that $\mathbb{P} \left( A_k = 1 \vert \overline{X}_k,  \overline{Y}_k = \overline{0}, \overline{Z}_k = \overline{0},  \overline{A}_{k-1} = \overline{a}_{k-1} \right) = 1$ if $\overline{a}_{k-1} \neq \overline{0}$,  because in our analysis we focus on the treatment initiation, hence once the treatment has been initiated at time $k$, i.e. $a_k = 1$, for some $k \in \left\lbrace 0, \dots, K-1 \right\rbrace$,  then the indicator for the observed treatment remains $1$ from that point onwards, i.e. $a_l = 1$, for $l \geq k$. Therefore, we only need to estimate these time--dependent propensity scores in patients in whom the treatment has not been initiated yet, which can be done by fitting a pooled logistic regression model
\begin{equation*}
\mathbb{P} \left( A_k = 1 \vert \overline{X}_k, \overline{Y}_k = \overline{0}, \overline{Z}_k = \overline{0}, \overline{A}_{k-1} = \overline{0} \right) = \expit \left\lbrace \gamma_0 + \gamma_1 k + \overline{\gamma_2}^\intercal \overline{X}_k \right\rbrace, \text{ for } k = 0, \dots, K-1.
\end{equation*}
\indent
In Table \ref{table1} the column "PS" contains values of propensity scores given by (\ref{eqn7}). Once we have the propensity scores we calculate the IPC weights ("IPCW") using expression (\ref{eqn6}). 

\subsection{Inverse Probability Weighted Aalen--Johansen Estimator} \label{subsection:label3}

In this section we let $V = \emptyset$ or, in other words, we consider the entire patient population. With interest in a specific subpopulation with characteristics $V$ one may perform exactly the same analysis restricting to the stratum $V = v$ (provided that it contains data on a sufficient number of individuals). Under the aforementioned assumptions, the counterfactual event--specific hazards $(\ref{eqn1})$ and $(\ref{eqn2})$ can be consistently estimated by
\begin{align}
\label{eqn8}
\hat{h}_{k}^{(1)} \left(d \right) &\equiv \frac{\sum_{i=1}^n Y_{k, i} \left(1 - Z_{k, i} \right) \left(1 - Y_{k-1, i} \right) \hat{W}_{d, k-1, i} }{\sum_{i=1}^n \left(1 - Z_{k, i} \right) \left(1 - Y_{k-1, i} \right) \hat{W}_{d, k-1, i} } \\
\label{eqn9}
\hat{h}_{k}^{(2)} \left(d \right) &\equiv \frac{\sum_{i=1}^n Z_{k, i} \left(1 - Y_{k-1, i} \right) \left(1 - Z_{k-1, i} \right) \hat{W}_{d, k-1, i}}{\sum_{i=1}^n \left(1 - Y_{k-1, i} \right) \left(1 - Z_{k-1, i} \right) \hat{W}_{d, k-1, i}}
\end{align}
where $\hat{W}_{d, k-1, i}$ is given by (\ref{eqn6}) with $k$ changed to $k-1$. Expressions ($\ref{eqn8}$) and ($\ref{eqn9}$) correspond to the inverse probability of censoring weighted estimators of the counterfactual event--specific hazards at time $k$ for the event of interest and for the competing event, respectively (see Lemma \ref{lemma1} in the Appendix \ref{app1}). Substituting expressions ($\ref{eqn8}$) and ($\ref{eqn9}$) into formula ($\ref{eqn3}$) we obtain the following estimator for the counterfactual cumulative incidence at time $k = 1, \dots, K$
\begin{equation}
\label{eqn10}
\hat{\mathbb{P}} \left( Y_k^d = 1 \right) = \sum_{l=1}^{k} \hat{h}_l^{(1)} \left(d \right) \left\lbrace 1 - \hat{h}_l^{(2)} \left(d \right) \right\rbrace \prod_{j=1}^{l-1} \left\lbrace 1-\hat{h}_j^{(1)} \left(d \right) \right\rbrace \left\lbrace 1-\hat{h}_j^{(2)} \left(d \right) \right\rbrace,
\end{equation}
which corresponds to an inverse probability of censoring weighted Aalen--Johansen estimator.
\\
\indent
Instead of modelling event--specific hazards (\ref{eqn8}) and (\ref{eqn9}), and then combining them into a cumulative incidence curve via the formula (\ref{eqn10}), one could aim at direct modelling the binary outcome $Y_k^d$ to obtain an IPC weighted estimator of $\mathbb{P} \left( Y_k^d = 1 \right)$ given by
\begin{equation*}
\hat{\mathbb{P}} \left( Y_k^d = 1 \right) = \frac{ \sum_{ i=1 }^{n} Y_{k,i} \hat{W}_{d, k-1, i} }{ \sum_{ i=1 }^{n} \hat{W}_{d, k-1, i} }.
\end{equation*}
The estimators that we consider are expected to have better performance. This can be seen by the fact that they generate monotone cumulative incidence curves. The survival curves produced through modelling $\mathbb{P} \left( Y_k^d = 1 \right)$ as a binary outcome consecutively for each $k \in \left\lbrace 1, \dots,K \right\rbrace$, would require performing IPC weighting estimation separately at each time point, and this could lead to non--monotonicity of the resulting survival curve resulting from the estimation uncertainty of the IPC weights. In such a case, additional techniques like isotonic regression \citep{Westling2020} would be required to restore the monotonicity of the cumulative incidence curves.  

\subsection{Dynamic--Regime Marginal Structural Models}

An alternative approach to estimate the cumulative incidence curve for different dynamic treatment regimes is to postulate dynamic--regime marginal structural models (DYR--MSM) \citep{Murphy2001, VanderLaan2007, Orellana2010} for the event--specific hazards ($\ref{eqn1}$) and ($\ref{eqn2}$). A DYR--MSM is a model for the counterfactual event--specific hazard of the event of interest, for instance, a pooled logistic model 
\begin{equation}
\label{eqn11}
h_k^{(1)} \left(d, V \right) = \expit \left\lbrace \alpha_0 + \alpha_1^\intercal d + \alpha_2 k + \alpha_3^\intercal d k + \alpha_4^\intercal V \right\rbrace \text{, for } k = 1, \dots, K,
\end{equation}
where $d$ is modelled as a categorical variable, i.e. it is a vector of dummy variables indicating the dynamic treatment regime. We denote the right hand side of (\ref{eqn11}) as $h_k^{(1)} \left(d, V; \alpha \right)$. Analogously, one may postulate a DYR--MSM for the counterfactual event--specific hazard of the competing event (ICU discharge), for instance
\begin{equation}
\label{eqn12}
h_k^{(2)} \left(d, V \right) = \expit \left\lbrace \beta_0 + \beta_1^\intercal d + \beta_2^\intercal k + \beta_3^\intercal d k + \beta_4^\intercal V \right\rbrace \text{, for } k = 1, \dots, K,
\end{equation}
where $d$ is again modelled as a categorical variable. We denote the right hand side of (\ref{eqn12}) as $h_k^{(2)} \left(d, V; \beta \right)$. Following the developments by \cite{Orellana2010}, and \cite{VanderLaan2007} we can consistently estimate the parameters indexing (\ref{eqn11}) and (\ref{eqn12}) by solving the following estimating equations 
\begin{equation}
\label{eqn13}
\sum_{i=1}^n \sum_{d \in \mathcal{D}} \sum_{k=1}^{K}  \left\lbrace Y_{k, i} - h_{k,i}^{(1)} \left(d, V; \alpha \right) \right\rbrace \left( 1 - Y_{k-1,i} \right) \left( 1 - Z_{k,i} \right) q_{k,i}^{(1)} \left(d, V \right) W_{d,k-1,i} = 0
\end{equation}
\begin{equation}
\label{eqn14}
\sum_{i=1}^n \sum_{d \in \mathcal{D}} \sum_{k=1}^{K} \left\lbrace Z_{k,i} - h_{k,i}^{(2)} \left(d, V; \beta \right) \right\rbrace \left( 1 - Z_{k-1,i} \right) \left( 1 - Y_{k-1,i} \right) q_{k,i}^{(2)}  \left(d, V \right) W_{d,k-1,i} = 0
\end{equation}
for $\alpha$ and $\beta$. See Lemma \ref{lemma3} in the Appendix \ref{app1} for technical details. Furthermore, choosing $q_k^{(1)} \left(d, V \right) = \left( 1, d, k, d k, V \right)^\intercal$ and $q_k^{(2)} \left(d, V \right) = \left( 1, d, k, d k, V \right)^\intercal$, the estimating equations (\ref{eqn13}) and (\ref{eqn14}) correspond to the weighted score equations of the following logistic regression models
\begin{equation}
\label{eqn15}
\mathbb{P} \left( Y_k = 1 \vert Y_{k-1} = 0, Z_k = 0, C_{d,k-1} = 1, V \right) = \expit \left\lbrace \tilde{\alpha}_0 + \tilde{\alpha}_1^\intercal d + \tilde{\alpha}_2^\intercal k + \tilde{\alpha}_3^\intercal d k + \tilde{\alpha}_4^\intercal V \right\rbrace
\end{equation}
\begin{equation}
\label{eqn16}
\mathbb{P} \left( Z_k = 1 \vert Z_{k-1} = 0, Y_{k-1} = 0, C_{d,k-1} = 1, V \right) =  \expit \left\lbrace \tilde{\beta}_0 + \tilde{\beta}_1^\intercal d + \tilde{\beta}_2^\intercal k  + \tilde{\beta}_3^\intercal d k + \tilde{\beta}_4^\intercal V \right\rbrace
\end{equation}
for $k = 1, \dots, K$ with weights given by (\ref{eqn6}). The resulting maximum weighted likelihood estimates of the parameters indexing (\ref{eqn15}) and (\ref{eqn16}) are consistent for the parameters indexing (\ref{eqn11}) and (\ref{eqn12}). Substituting the event--specific hazards $h_k^{(1)} \left(d, V; \hat{\alpha} \right)$ and $h_k^{(2)} \left(d, V; \hat{\beta} \right)$ in formula $(\ref{eqn3})$ we obtain the following consistent estimator of the cumulative incidence at time $k = 1, \dots, K$ under regime $d$ in the subpopulation $V$
\begin{align}
\label{eqn17}
&\hat{\mathbb{P}} \left( Y_k^d = 1 \vert V \right) \\
\nonumber
&= \sum_{l=1}^{k} h_l^{(1)} \left(d, V; \hat{\alpha} \right) \left\lbrace 1 - h_l^{(2)} \left(d, V; \hat{\beta} \right) \right\rbrace \prod_{j=1}^{l-1} \left\lbrace 1-h_j^{(1)} \left(d, V; \hat{\alpha} \right) \right\rbrace \left\lbrace 1-h_j^{(2)} \left(d, V; \hat{\beta} \right) \right\rbrace.
\end{align}
By repeating this calculation for each time $k$, this allows us to compute the cumulative incidence curve for each of the dynamic treatment regimes. Furthermore, by averaging over $V$, i.e. obtaining a curve for each patient given his/her observed covariates $V$ and then averaging across all patients, we obtain the marginal cumulative incidence curve for the entire population. 
\\
\indent
In the example introduced in the section \ref{subsection:label2}, to estimate parameters $\alpha$ and $\beta$ we fit models (\ref{eqn15}) and (\ref{eqn16}) to the extended data set restricted to rows with compatibility indicator equal to 1 on the previous day for the patients who are still hospitalized at the intensive care unit, applying the IPC weights given by (\ref{eqn6}) to account for artificial censoring due to incompatibility with the given dynamic treatment regime.

\subsubsection{DYR--MSM for Heterogeneous Treatment Effects}\label{subsection:label1}

The effect of conditioning on $V$ in the marginal structural model is twofold. First, by conditioning on the set of covariates $V$ we permit a different optimal decision rule across different strata $V$, even if regime $d$ is not formally defined as a function of $V$. Indeed, imagine we want to compare dynamic treatment regimes $d^{(1)}$, $d^{(2)}$ and $d^{(3)}$. If we choose $V = \emptyset$ then the optimal rule is one of the three considered rules applied on the whole population. However, if we condition on $V$ then the optimal rule might vary between different strata of the population defined by different levels of $V$, e.g. $d^{(3)}$ applied to patients with $V = v_1$ and $d^{(1)}$ applied to patients with $V = v_2$. Therefore, conditional DYR--MSMs allow searching for the optimal rule over a bigger set of rules that, in addition to the aforementioned time--updated physiological parameters, also take into account baseline patient characteristics. Second, conditioning on a set of covariates $V$ in the MSM permits obtaining more stable estimates by applying stabilized IPC weights. If we use stabilized weights $SW_{d,k}$ (to be defined below) we need to make sure that the resulting estimating equations 
\begin{equation}
\label{eqn18}
\sum_{i=1}^n \sum_{d \in \mathcal{D}} \sum_{k=1}^{K} \left\lbrace Y_{k, i} - h_{k,i}^{(1)} \left(d, V; \alpha \right) \right\rbrace \left( 1 - Y_{k-1,i} \right) \left( 1 - Z_{k,i} \right) q_{k,i}^{(1)} \left(d, V \right) SW_{d,k-1,i} = 0
\end{equation}
\begin{equation}
\label{eqn19}
\sum_{i=1}^n \sum_{d \in \mathcal{D}} \sum_{k=1}^{K} \left\lbrace Z_{k,i} - h_{k,i}^{(2)} \left(d, V; \beta \right) \right\rbrace \left( 1 - Z_{k-1,i} \right) \left( 1 - Y_{k-1,i} \right) q_{k,i}^{(2)} \left(d, V \right) SW_{d,k-1,i} = 0
\end{equation}
remain unbiased. Following Cain et al.  \cite{Cain2010}, we argue that the numerator of the stabilized weight $SW_{d,k}$ can only depend on $d$ and $V$, hence $SW_{d,k} \equiv f_{k} \left( d, V \right) \times W_{d,k}$ for some function $f_{k} \left( d, V \right)$. If the numerator $f_{k} \left( d, V \right)$ is only a function of $d$ and $V$, then the unbiasedness of the estimating equations (\ref{eqn18}) and (\ref{eqn19}) follows from the fact that the estimating equations (\ref{eqn13}) and (\ref{eqn14}) are unbiased for all $q_{k,i}^{(j)} \left( d, V \right)$ for $j=1,2$. On the other hand, if the numerator $f_{k}$ of the stabilized weight $SW_{d,k}$ depends on other variables than $d$ and $V$, then equations (\ref{eqn18}) and (\ref{eqn19}) are no longer unbiased for all $q_{k,i}^{(j)} \left( d, V \right)$ for $j=1,2$.
Therefore, the numerator of the stabilized IPC weights $f_{k}$ can only depend on the treatment regime $d$ and the set of baseline covariates $V$. In this case a natural choice for the numerator of the stabilized IPC weights is 
\begin{align*}
f_{k} \left( d, V \right) &\equiv \prod_{l=1}^{k} \mathbb{P} \left\lbrace A_{l} = d_l \left( \overline{O}_{l}, \overline{A}_{l-1} \right) \vert \overline{Y}_{l} = \overline{0}, \overline{Z}_{l} = \overline{0}, C_{d,l-1} = 1, V \right\rbrace.
\end{align*}
Indeed, this choice ensures that when
\begin{equation*}
A_l \cip \overline{O}_l \vert \overline{Z}_{l-1}, \overline{Y}_{l-1}, \overline{A}_{l-1}, V
\end{equation*}
i.e. the confounding is solely due to baseline confounders or previous exposures, then the resulting estimator is efficient since the weights $SW_{d,k} = 1$ for all $k = 1, \dots, K$ and therefore the resulting estimator is just the maximum likelihood estimator and hence is efficient.

\section{Empirical Example}\label{sec5}

\subsection{Data}

The ICIS (Intensive Care Information System) of the Ghent University Hospital ICUs contains records from all adult patients admitted to the intensive care unit since $2013$. Our analysis is restricted to $4769$ retrospectively identified stage $2$ AKI patients \citep{Vanmassenhove2021} (based on the KDIGO -- Kidney Disease: Improving Global Outcomes -- criteria \citep{KDIGO2012}), admitted between $1/1/2013$ and $31/12/2017$ and who satisfied the following criteria by the time of the inclusion at stage $2$ AKI diagnosis: no recorded RRT history, aged $18$ years or above, no stage $3$ AKI diagnosis and no dialysis restriction. 
\\
\indent
For each patient we have information on ICU admission time, ICU discharge time, vital status at discharge, timestamps of all dialysis sessions during each ICU episode, baseline covariates (e.g. age, weight, gender, admission category \{"No surgery", "Planned surgery", "Emergency surgery"\}, indicator for whether patient has received dialysis prior to current ICU admission, indicator for whether patient had chronic kidney disease diagnosis prior to current ICU admission) and longitudinal measurements (e.g. SOFA scores, indicator for whether KDIGO AKI (stage $1/2/3$) creatinine condition has been reached during ICU episode, indicator for whether KDIGO AKI (stage $1/2/3$) oliguric condition has been reached during ICU episode, cumulative total fluid intake, cumulative total fluid output, arterial pH values, serum potassium values (in mmol/L), serum ureum values (in mg/dL), serum magnesium values (in mmol/L), fraction of inspired oxygen (FiO$_2$), peripheral oxygen saturation (SpO$_2$), arterial oxygen concentration (PaO$_2$), ratio of arterial oxygen concentration to the fraction of inspired oxygen (P/F ratio), DNR ("Do Not Resuscitate") code) and their timestamps. 

\subsection{Results}

\begin{figure}[H]
\begin{center}
\includegraphics[scale=0.75]{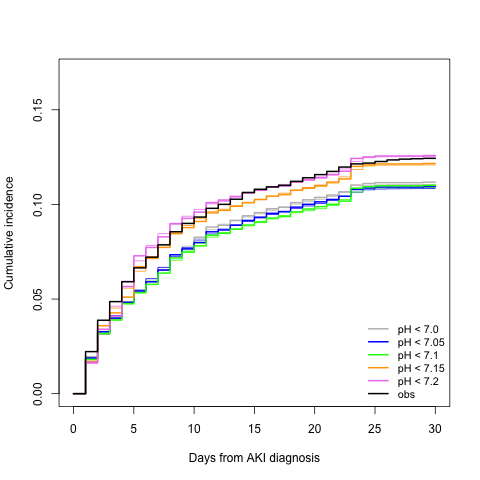}
\end{center}
\caption{Cumulative incidence curves for considered DTRs (pH $< x$) and observed cumulative incidence under the current standard of care (obs). Thick lines correspond to the estimates from the Aalen--Johansen estimator and thin lines correspond to the estimates from the dynamic--regime marginal structural model.\label{figure1}}
\end{figure}

\begin{figure}[H]
\begin{center}
\includegraphics[scale=0.75]{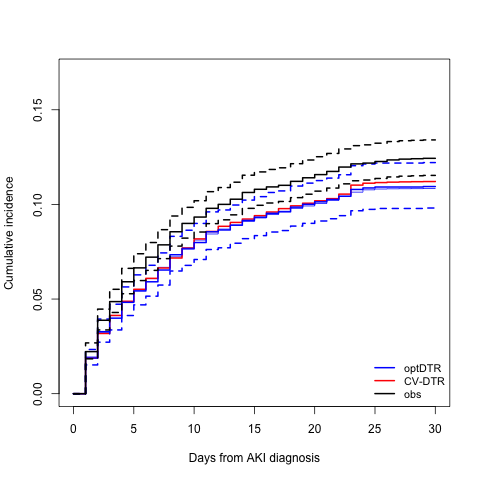}
\end{center}
\caption{Cumulative incidence curves for the optimal DTR (optDTR), $5$--fold cross--validation cumulative incidence curve for the optimal DTR (CV--DTR) and the current standard of care (obs). The dashed lines correspond to $95\%$ confidence intervals for the Aalen--Johansen estimators.\label{figure2}}
\end{figure}

\begin{figure}[H]
\begin{center}
\includegraphics[scale=0.52]{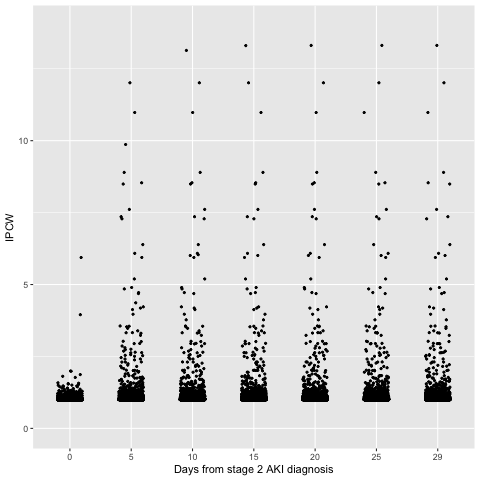}
\includegraphics[scale=0.52]{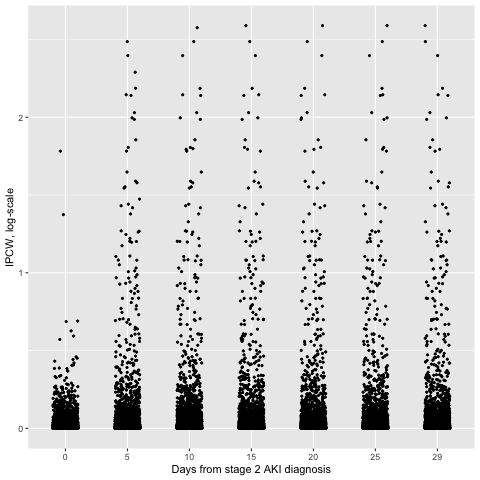}
\end{center}
\caption{IPC weights for the estimated optimal DTR.\label{figure3}}
\end{figure}

We consider dynamic treatment regimes that initiate RRT if a patient meets the AKI Stage $2$ creatinine criterion and in the last $24$h patient's pH fell below threshold $x$ or the oliguric stage $3$ AKI criterion had been met. We investigated different pH thresholds $x \in \left\lbrace 7.0, 7.05, 7.1, 7.15, 7.2 \right\rbrace$ in order to identify the treatment rule resulting in the lowest $30$--day ICU mortality.  
\\
\indent
To perform our analysis we have constructed an extended data set analogous to the example presented in the Table \ref{table1}. This means we have a copy of the original data for each of the dynamic treatment regimes corresponding to different pH thresholds. Patients are followed up from the time when the stage $2$ AKI criteria are satisfied for the first time (day $0$) until ICU death or discharge on day $30$, whichever occurs first. We assume that the decision about RRT initiation based on the considered DTRs is performed every $24$h starting from day $0$ based on the information available up to a given decision point. If the treatment initiation decision based on the considered DTR does not coincide with the observed treatment, then the patient is censored for this particular DTR from this time point onward, as described in the example in Section \ref{subsection:label2}. We applied IPC weighting to deal with potential selection bias due to this artificial censoring. Once we have constructed the extended data set we calculate the cumulative incidence under each prespecified DTR using the IPC weighted Aalen--Johansen estimator given by (\ref{eqn10}) and marginal structural models given by (\ref{eqn17}) with $V = \emptyset$, which means that we do not include any baseline covariates in the DYR--MSMs for the counterfactual event--specific hazards. We have made this choice in our empirical example since the interest of clinicians was in simple rules that do not change across different levels of baseline covariates. Moreover we have investigated the use of stabilized weights in our analysis, but this did not lead to improvement in the stability of weights. Therefore we used unstabilized IPC weights in the final analysis. The resulting MSM--based estimates for the cumulative incidence curve should thus approximate the corresponding weighted Aalen--Johansen estimator. We consider the optimal DTR to be the one with the lowest $30$--day ICU mortality. Figure \ref{figure1} presents the counterfactual cumulative incidence curves under different considered DTRs and the observed cumulative incidence under current standard of care (obs). 
\indent
Figure \ref{figure2} presents the cumulative incidence curves for the optimal dynamic treatment regime "initiate RRT if creatinine stage $2$ AKI condition has been met and one of the events occurred in the last $24$h: pH $< 7.05$ or oliguric stage $3$ AKI condition has been met", which resulted in the lowest estimated $30$--day ICU mortality among the considered DTRs, computed using the Aalen--Johansen estimator given by (\ref{eqn10}) and the estimator obtained using marginal structural models given by (\ref{eqn17}). The suggested treatment rule leads to a $12.6 \%$ relative decrease in $30$--day ICU mortality from $12.43 \%$ for the current standard of care to $10.85 \%$. The confidence intervals for the IPC weighted Aalen--Johansen estimator are based on the infinitesimal jackknife variance estimator, as implemented in the package \texttt{survival} \citep{Therneau2020}, and are presented by dashed lines. They are expected to be conservative as a result of ignoring uncertainty in the obtained weights \citep{Rotnitzky2010a}. The confidence intervals for the marginal structural model estimator can be obtained using the nonparametric bootstrap. These are not displayed in Figure \ref{figure2}, as these nearly coincide with confidence intervals for the IPC weighted Aalen--Johansen estimator.  Note that these confidence intervals have been computed for a particular prespecified treatment regime and as such do not take into account the uncertainty due to the choice of the optimal treatment regime (see Section \ref{sec6} for further discussion of this issue).
\\
\indent
As indicated in Section \ref{sec3} there is a risk of being overly optimistic about the optimal DTR whenever its selection and its performance assessment are conducted on the same data set. To provide a more realistic out--of--sample performance assessment we computed using $5$--fold cross--validation the benchmark for the optimal dynamic treatment regime given by (\ref{eqn4}). An important question when combining the IPC weights and cross--validation is whether to re--estimate the weights on the training set for each fold. We have chosen to estimate the weights on the complete data prior to performing cross--validation in order to prevent fitting the propensity score on smaller sample sizes. We can see that the cross--validation estimate of the mortality risk (CV--DTR) results in a reduction of $30$--day ICU mortality compared to the current standard of care, however the reduction is slightly smaller than suggested by the in--sample estimate. The difference between the cross--validation benchmark (CV--DTR) and the in--sample evaluation of $30$--day ICU mortality under the optimal DTR (optDTR) in Figure \ref{figure2} gives an indication of over--optimism due to choice and performance assessment of the optimal regime on the same data set. There thus remains a suggestion that the estimated optimal treatment regime is expected to lead to a reduction in ICU mortality relative to the standard of care. 
\indent
Figure \ref{figure3} presents the unstabilized IPC weights on a number of selected days that have been used to compute the cumulative incidence curves for the best performing dynamic treatment regime. We present plots of the IPC weights on the linear scale in the upper plot and on the logarithmic scale in the lower plot. 
\indent
Additionally, we investigated the proportion of stage $2$ AKI patients who would receive RRT under the best performing DTR and compared it to current practice. This is relevant from a clinical perspective because potentially we could have found a treatment rule that significantly improves ICU survival, but leads to such a high number of additional patients in need of RRT that it would become logistically impossible to provide treatment to all. In Figure \ref{figure4} we present the comparison of the proportion of stage $2$ AKI patients that would receive RRT under the suggested treatment rule (optDTR), which is estimated by the sum of (unstabilized) IPC weights of the treated compatible patients divided by the sum of IPC weights of all compatible patients at a given time point. This follows from the identity
\begin{equation*}
\mathbb{E} \left( A_{k}^d \right) = \frac{\mathbb{E} \left( A_{k} W_{d,k-1} \right)}{\mathbb{E} \left( W_{d,k-1} \right)}
\end{equation*}
where $A_k^d$ is the potential outcome for treatment under regime $d$, which can be shown following a similar calculation as in the proof of Lemma $\ref{lemma1}$ (see the Appendix \ref{app1}). Based on this estimate, a smaller proportion of patients would receive RRT under the optimal DTR as compared to the current standard of care. 
\begin{figure}[H]
\begin{center}
\includegraphics[scale=0.75]{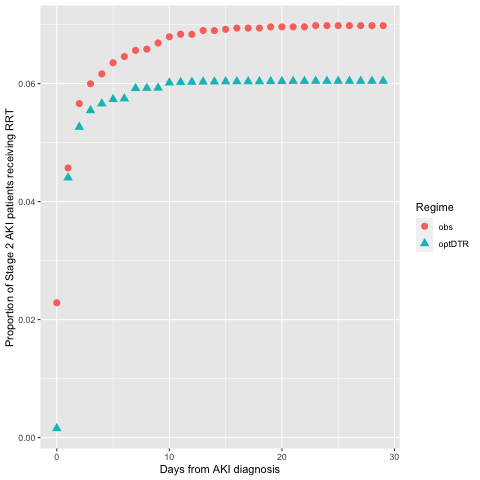}
\end{center}
\caption{Proportion of stage $2$ AKI patients who would receive RRT under the best performing among considered DTRs (optDTR) and proportion of stage $2$ AKI patients who receive RRT under the current standard of care (obs).\label{figure4}}
\label{figure4}
\end{figure}

\section{Discussion}\label{sec6}

In this paper, we illustrated how routinely collected data can be leveraged to inform decisions about RRT initiation among future AKI patients in an intensive care unit. For this purpose we considered two estimators -- an inverse probability weighted nonparametric Aalen--Johansen estimator and an estimator based on semiparametric dynamic--regime marginal structural models -- to deal with the problem of competing risks and time--varying treatment and confounding. Furthermore, we proposed a cross--validation benchmark to compare the performance of the in--sample selected optimal DTR against, and hence to assess the amount of over--optimism that characterizes the in--sample performance assessment of this optimal DTR.  
\\
\indent
To the best of our knowledge \cite{VanderLaan2014} were the first ones to use cross--validation in context of dynamic treatment regimes. These authors apply cross--validation in order to perform statistical inference for the mean outcome under the estimated optimal dynamic treatment regime.  In contrast, the purpose of cross--validation in our analysis is to compute a cross--validation benchmark for developing insight into the extent to which a different regime may appear optimal on a different subsample, and to obtain a finite--sample evaluation of the degree of optimism due to estimating the optimal regime and performance under that regime from the same sample. This evaluation is less ambitious, and therefore also simpler, than in \cite{VanderLaan2014} and \cite{Andrews2018}, who obtain confidence intervals for the mean outcome under the estimated optimal treatment regime that take into account uncertainty related to the choice of the optimal regime.  Such less ambitious evaluations are also common when studying the predictive performance of statistical models, where they have become mainstream. We hope that the simplicity of our proposal will likewise encourage its uptake and thereby prevent the winner's curse \citep{Andrews2018}. 
\\
\indent
One of the limitations of our work is the focus on ICU mortality, instead of all--cause mortality. However, we believe this not to induce major concerns in the considered setting since the chance of dying shortly after ICU discharge can be expected to be low, and therefore ICU death to approximate well all--cause mortality in the short term. However, in a general competing events setting, our proposed approach might pose additional challenges.  For instance, if the chance of death after ICU discharge were not low, and therefore ICU death would not approximate well all--cause mortality in the short term,  then modelling the impact of a given treatment regime on ICU mortality and on ICU discharge could pose interpretative difficulties.  For instance, imagine we would like to compare two treatment regimes $d^{(1)}$ and $d^{(2)}$. Furthermore, assume that their impact on mortality is the same, however treatment $d^{(1)}$ leads to longer stays at the ICU. Now, merely due to longer ICU stays we may observe more ICU deaths under treatment regime $d^{(1)}$, and therefore treatment regime $d^{(2)}$ may appear beneficial, despite the fact that the impact of both treatment regimes on death is the same.  As a remedy, following the lines of \cite{Gran2015}, one could allow the treatment regime to alter only some of the event--specific hazards, while keeping the remaining ones unchanged.  This means one could model the counterfactual hazard for the event of interest under the considered treatment regime the same way as we did in our analysis, however fixing the hazard for the competing event at that obtained under the current standard of care, therefore assuming no impact of DTR on the competing event mechanism. In such analysis one would be able to compare different DTRs solely based on their impact on the event of interest mechanism and not through potential impact on the competing event mechanism. However, such analysis would necessitate control for all risk factors of ICU discharge that are also associated with survival, which would be difficult to attain given that these processes are so intimately linked.
\\
\indent
Another limitation of our analysis is the fact that we rely on correct specification of parametric models -- the propensity score model and possibly marginal structural hazard models. One way to avoid these modelling demands would be to use data--adaptive methods via longitudinal minimum loss--based estimation for the $30$--day ICU counterfactual mortality \citep{Petersen2014, Lendle2017}. 
\\
\indent
A final limitation is that medical doctors may be hesitant to base their decision--making on simplified decision rules,  which take into account only a limited set of available information.  They may arguably be better served with a decision support model,  which computes individual patient's survival chances in either scenario of treatment initiation or its delay.  In future work we therefore plan to focus on sequential counterfactual prediction \citep{Wager2018, Coston2020},  which could be used as a framework for such decision support. This permits to incorporate more flexible data--adaptive algorithms into the procedure and to consider treatment rules that are much more tailored to individual patient characteristics. Since the final decision would still rely on the medical doctor, they could incorporate additional information not available to the decision--support model into their decision--making process, which often takes multiple parameters into account in a complex fashion. 

\section*{Acknowledgments}
Paweł Morzywołek,  Johan Steen,  Wim Van Biesen and Stijn Vansteelandt were supported by the Flemish Research Council (FWO Research Project 3G068619 — Grant FWO.OPR.2019.0045.01). The authors would like to thank Bram Gadeyne, Veerle Brams and Christian Danneels for technical support.

\section*{Data Availability Statement}

The data are not publicly available due to confidentiality restrictions.

\section*{Supporting information}

Additional supporting information, including extension of the proposed methodology to allow for censoring due to administrative end of study or loss to follow--up and proofs of the main results, may be found in the Appendix. The R code to reproduce our data analysis on synthetic data from the package \texttt{ltmle} \citep{Lendle2017} is available at \texttt{https://github.com/pmorzywolek/comprisksDTR}. The data set has been modified to include also a competing event (see data preparation part in the R code).

\bibliography{DynamicTreatmentRegimes2}
\bibliographystyle{apalike}

\newpage

\appendix
\section*{Appendix}

\section{Censoring due to administrative end or loss to follow-up } \label{app0}

In the main text we assume no censoring due to administrative end or loss to follow-up, since the focus of our methodology is to deal with artificial censoring due to incompatibility of an individual with a dynamic treatment regime whose effect we wish to infer in the setting with competing risks. Below we discuss the extensions, including the additional assumption on the censoring mechanism and the updated version of IPC-weighting, which are required to allow for censoring due to administrative end or loss to follow-up in the proposed methodology. 
\\
\indent
Let $U_k$, for $k=1,\dots,K$, indicate censoring due to end or loss to follow-up (with $U_k = 0$ if an individual has been censored by time $k$ and $U_k = 1$ otherwise). We assume the following order of the variables within the $k^{th}$ time interval $\left( Z_k, Y_k, X_k, A_k, U_k \right)$. The ordering needs to be aligned with our understanding of the data-generating mechanism and consistently applied across time intervals.  For all patients in the study population $U_k = 1$ for $k \leq 0$.  

\subsection{Identification of the Causal Estimands}

Allowing for censoring due to administrative end or loss to follow-up in our analysis we need to introduce also an assumption on the missingness mechanism to identify the causal estimand $\mathbb{P} \left( Y_k^d = 1 \vert V = v \right)$, for $k = 1,\dots, K$. The required assumption on the missingness mechanism is given by
\begin{equation} \label{eqnA1}
U_k \cip \left( \underline{Z}_{k+1}, \underline{Y}_{k+1}, \underline{X}_{k+1}, \underline{A}_{k+1} \right) \vert \overline{U}_{k-1}, = \overline{1},  \overline{Z}_{k} = \overline{0}, \overline{Y}_{k} = \overline{0}, \overline{X}_{k}, \overline{A}_{k}, \text{ for } k = 1, \dots, K,
\end{equation}
i.e. the censoring indicator $U_k$ is assumed to be independent of the future values \\
$\left( \underline{Z}_{k+1}, \underline{Y}_{k+1}, \underline{X}_{k+1}, \underline{A}_{k+1} \right)$ given the past (i.e. $\overline{U}_{k-1}, = \overline{1},  \overline{Z}_{k} = \overline{0}, \overline{Y}_{k} = \overline{0}, \overline{X}_{k}, \overline{A}_{k}$) in the population that is uncensored and at risk of ICU death or discharge by the time $k$.

\subsection{Inverse Probability of Censoring Weighting}

Allowing for censoring due to end or loss to follow-up in our analysis we need to deal with two types of censoring, i.e. censoring due to end or loss of follow-up and artificial censoring due to incompatibility with the considered treatment regime. Artificial censoring due to incompatibility with a treatment regime $d$ at time $k$ might select a group of patients that is no longer comparable with the risk set under the observed regime in terms of prognostic factors for the outcome. One approach to remedy selection bias due to this artificial censoring is the use of inverse probability of censoring weighting (IPC-weighting). IPC weights are given by
\begin{equation*}
W_{d,k,i} \equiv \frac{C_{d,k,i}}{ \prod_{l=1}^k \pi^C_{d,l,i}} \times \frac{\prod_{l=1}^{k} U_{l,i}}{ \prod_{l=1}^k \pi^U_{l,i}},
\end{equation*}
where $\pi^C_{d,l,i}$ is the probability that an individual $i$ remains compatible with the DTR $d$ by the time $l$ conditional on the observed history and remaining compatible with the DTR $d$ and uncensored until the time $l-1$, given by 
\begin{align} \label{eqnA2}
&\pi^C_{d,l,i} \equiv \mathbb{P} \left\lbrace C_{d,l,i} = 1 \vert \overline{U}_{l,i} = \overline{1}, \overline{O}_{l,i}, C_{d,l-1,i} = 1 \right\rbrace \\
\nonumber
&=
\begin{cases}
\mathbb{P} \left\lbrace C_{d,l,i} = 1 \vert \overline{U}_{l,i} = \overline{1},  \overline{Y}_{l,i} = \overline{0}, \overline{Z}_{l,i} = \overline{0}, \overline{X}_{l,i},  C_{d,l-1,i} = 1 \right\rbrace, \text{if } \overline{U}_{l-1,i} = \overline{1},  \overline{Y}_{l,i} = \overline{0} \text{ and } \overline{Z}_{l,i} = \overline{0} \\
1, \text{ otherwise}
\end{cases}
\end{align}
and $\pi^U_{l,i}$ is the probability that an individual $i$ remains uncensored by the time $l$ conditional on the observed history and remaining compatible with the DTR $d$ and uncensored until the time $l-1$, i.e.
\begin{align*} 
\nonumber
&\pi^U_{l,i} \equiv \mathbb{P} \left\lbrace U_{l,i} = 1 \vert \overline{U}_{l-1,i}, \overline{O}_{l,i}, \overline{A}_{l,i} \right\rbrace \\
&= 
\begin{cases}
\mathbb{P} \left( U_{l,i} = 1 \vert \overline{U}_{l-1,i} = \overline{1}, \overline{Y}_{l,i} = \overline{0}, \overline{Z}_{l,i} = \overline{0}, \overline{X}_{l,i}, \overline{A}_{l,i} \right),\text{ if } \overline{U}_{l-1,i} = \overline{1},  \overline{Y}_{l,i} = \overline{0} \text{ and } \overline{Z}_{l,i} = \overline{0} \\
1,\text{ otherwise.}
\end{cases}
\end{align*}
Note that $\pi^C_{d,l}$ differs from $\pi_{d,l}$ in the main text as it also includes conditioning on the censoring indicator, however under the missingness assumption (\ref{eqnA1}) we have $\pi^C_{d,l} = \pi_{d,l}$ for all $l$ and $d$. To obtain the estimates $\hat{\pi}^C_{d,l,i}$, we note first that the upper part of the expression (\ref{eqnA2}) can be written as
\begin{align*}
&\mathbb{P} \left\lbrace A_{l,i} = d_{l,i} \left( \overline{O}_{l,i}, \overline{A}_{l-1,i} \right) \vert \overline{U}_{l,i} = \overline{1}, \overline{Y}_{l,i} = \overline{0}, \overline{Z}_{l,i} = \overline{0}, \overline{X}_{l,i},  C_{d,l-1,i} = 1 \right\rbrace \\
&=\mathbb{P} \left\lbrace A_{l,i} = 1 \vert \overline{U}_{l,i} = \overline{1},  \overline{Y}_{l,i} = \overline{0}, \overline{Z}_{l,i} = \overline{0}, \overline{X}_{l,i},  C_{d,l-1,i} = 1 \right\rbrace^{d_{l,i} \left( \overline{O}_{l,i}, \overline{A}_{l-1,i} \right)} \\
&\times \left[ 1-\mathbb{P} \left\lbrace A_{l,i} = 1 \vert \overline{U}_{l,i} = \overline{1},  \overline{Y}_{l,i} = \overline{0}, \overline{Z}_{l,i} = \overline{0}, \overline{X}_{l,i},  C_{d,l-1,i} = 1 \right\rbrace \right]^{1-d_{l,i} \left( \overline{O}_{l,i}, \overline{A}_{l-1,i} \right)}.
\end{align*}
Furthermore, note that for individuals compatible with the DTR $d$ by the time $l$ we have
\begin{align*}
& \mathbb{P} \left( A_{l,i} = 1 \vert \overline{U}_{l,i} = \overline{1},  \overline{Y}_{l,i} = \overline{0}, \overline{Z}_{l,i} = \overline{0}, \overline{X}_{l,i},  C_{d,l-1,i} = 1 \right) \\
&=
\begin{cases}
\mathbb{P} \left( A_{l,i} = 1 \vert \overline{U}_{l,i} = \overline{1},  \overline{Y}_{l,i} = \overline{0}, \overline{Z}_{l,i} = \overline{0},  \overline{X}_{l,i},  \overline{A}_{l-1,i} = \overline{a}_{l-1,i} \right),\text{ if } \overline{a}_{l-1,i} = \overline{0} \\
1,\text{ otherwise}
\end{cases}
\end{align*}
where $\mathbb{P} \left( A_{l,i} = 1 \vert \overline{U}_{l,i} = \overline{1},  \overline{Y}_{l,i} = \overline{0}, \overline{Z}_{l,i} = \overline{0}, \overline{X}_{l,i},\overline{A}_{l-1,i} = \overline{a}_{l-1,i} \right)$ is the time-dependent propensity score, i.e. the probability of receiving treatment at a given time-point given the observed event and covariate history. Because $\mathbb{P} \left( A_{l,i} = 1 \vert \overline{U}_{l,i} = \overline{1},  \overline{Y}_{l,i} = \overline{0}, \overline{Z}_{l,i} = \overline{0},  \overline{X}_{l,i},  \overline{A}_{l-1,i} = \overline{a}_{l-1,i} \right) = 1$ if $\overline{a}_{l-1} \neq \overline{0}$,  we only need to estimate time-dependent propensity scores in patients in whom the treatment has not been initiated yet, which can be done by fitting a pooled logistic regression model
\begin{equation*}
\mathbb{P} \left( A_l = 1 \vert \overline{U}_{l} = \overline{1}, \overline{Y}_l = \overline{0}, \overline{Z}_l = \overline{0}, \overline{X}_l,  \overline{A}_{l-1} = \overline{0} \right) = \expit \left\lbrace \gamma_0 + \gamma_1 l + \overline{\gamma_2}^\intercal \overline{X}_k \right\rbrace, \text{ for } k = 0, \dots, K-1.
\end{equation*}
Similarly,  to obtain the estimates $\hat{\pi}^U_{l,i}$,  we fit a pooled logistic regression model
\begin{equation*}
\mathbb{P} \left( U_l = 1 \vert \overline{U}_{l-1} = \overline{1}, \overline{Y}_{l} = \overline{0}, \overline{Z}_{l} = \overline{0}, \overline{X}_{l}, \overline{A}_{l} \right) = \expit \left\lbrace \delta_0 + \delta_1 l + \overline{\delta_2}^\intercal \overline{X}_{l} + \overline{\delta_3}^\intercal \overline{A}_{l} \right\rbrace, \text{ for } l = 0, \dots, K-1.
\end{equation*}

\subsection{Stabilized IPC Weights}

The numerator of the stabilized IPC weights $f_{k}$ is only allowed to depend on the treatment regime $d$ and the set of baseline covariates $V$. In this case a natural choice for the numerator of the stabilized IPC weights is 
\begin{align*}
f_{k} \left( d, V \right) &\equiv \prod_{l=1}^{k} \mathbb{P} \left\lbrace A_{l} = d_l \left( \overline{O}_{l}, \overline{A}_{l-1} \right) \vert \overline{U}_{l-1} = \overline{1}, \overline{Y}_{l} = \overline{0}, \overline{Z}_{l} = \overline{0}, C_{d,l-1} = 1, V \right\rbrace \\
&\times \prod_{j=1}^{k} \mathbb{P} \left\lbrace U_{j} = 1 \vert \overline{U}_{j-1} = \overline{1}, \overline{Y}_{j} = \overline{0}, \overline{Z}_{j} = \overline{0}, \overline{A}_{j}, V \right\rbrace.
\end{align*}
Indeed, this choice ensures that when
\begin{equation*}
A_l \cip \overline{O}_l \vert \overline{Z}_{l-1}, \overline{Y}_{l-1}, \overline{A}_{l-1}, \overline{U}_{l-1}, V
\end{equation*}
i.e. the confounding is solely due to baseline confounders, then the resulting estimator is efficient since the weights $SW_{d,k} = 1$ for all $k = 1, \dots, K$ and therefore the resulting estimator is just the maximum likelihood estimator and hence is efficient.

\section{Lemmas} \label{app1}

\begin{lemma}
\label{lemma1}
Under the positivity, consistency and sequential randomization assumptions and assuming no model misspecification for the propensity score model in (\ref{eqn6}) we have $\hat{h}_k^{(1)}$ and $\hat{h}_k^{(2)}$ are consistent estimators of $h_k^{(1)}$ and $h_k^{(2)}$. 
\end{lemma}

\begin{proof}[Proof of Lemma~\ref{lemma1}]
We prove Lemma~\ref{lemma1} in the general case allowing for censoring due to administrative end or loss to follow-up. The version of the Lemma~\ref{lemma1} without censoring (due to administrative end or loss to follow-up) follows as a special case. In particular, note that the proof is structured into two parts, i.e. first dealing with censoring (due to administrative end or loss to follow-up) and second dealing with incompatibility with respect to a particular DTR (starting from (\ref{eqnB2})).   \\
We have the following
\begin{align*}
&\frac{\mathbb{E} \left\lbrace Y_{k} \left(1-Z_{k} \right) \left(1-Y_{k-1} \right) W_{d,k-1} \right\rbrace}{\mathbb{E} \left\lbrace \left(1-Z_{k} \right) \left(1-Y_{k-1} \right) W_{d,k-1} \right\rbrace} = \frac{\mathbb{E} \left\lbrace \frac{C_{d,k-1}}{\prod_{l=1}^{k-1}  \pi^C_{d,l}} \frac{ \prod_{l=1}^{k-1} U_{l}}{\prod_{l=1}^{k-1}  \pi^U_{l}} Y_{k} \left(1-Z_{k} \right) \left(1-Y_{k-1} \right) \right\rbrace}{\mathbb{E} \left\lbrace \frac{C_{d,k-1}}{\prod_{l=1}^{k-1}  \pi^C_{d,l}} \frac{ \prod_{l=1}^{k-1} U_{l}}{\prod_{l=1}^{k-1}  \pi^U_{l}} \left(1-Z_{k} \right) \left(1-Y_{k-1} \right) \right\rbrace} 
\end{align*}
Applying the law of iterated expectation
\begin{align*}
&= \frac{\mathbb{E} \left[ \mathbb{E} \left\lbrace \frac{C_{d,k-1}}{\prod_{l=1}^{k-1}  \pi^C_{d,l}} \frac{ \prod_{l=1}^{k-1} U_{l}}{\prod_{l=1}^{k-1}  \pi^U_{l}} Y_{k} \left(1-Z_{k} \right) \left(1-Y_{k-1} \right) \vert \overline{U}_{k-2}, \overline{O}_{k-1}, \overline{A}_{k-1}, \underline{O}_{k}, \underline{A}_{k} \right\rbrace \right]}{\mathbb{E} \left[ \mathbb{E} \left\lbrace \frac{C_{d,k-1}}{\prod_{l=1}^{k-1}  \pi^C_{d,l}} \frac{ \prod_{l=1}^{k-1} U_{l}}{\prod_{l=1}^{k-1}  \pi^U_{l}} \left(1-Z_{k} \right) \left(1-Y_{k-1} \right) \vert  \overline{U}_{k-2}, \overline{O}_{k-1}, \overline{A}_{k-1}, \underline{O}_{k}, \underline{A}_{k} \right\rbrace \right]} \\
&= \frac{\mathbb{E} \left[ \frac{C_{d,k-1}}{\prod_{l=1}^{k-1}  \pi^C_{d,l}} \frac{ \prod_{l=1}^{k-2} U_{l}}{\prod_{l=1}^{k-1}  \pi^U_{l}} Y_{k} \left(1-Z_{k} \right) \left(1-Y_{k-1} \right) \mathbb{E} \left\lbrace U_{k-1} \vert \overline{U}_{k-2}, \overline{O}_{k-1}, \overline{A}_{k-1}, \underline{O}_{k}, \underline{A}_{k} \right\rbrace \right]}{\mathbb{E} \left[  \frac{C_{d,k-1}}{\prod_{l=1}^{k-1}  \pi^C_{d,l}} \frac{ \prod_{l=1}^{k-2} U_{l}}{\prod_{l=1}^{k-1}  \pi^U_{l}} \left(1-Z_{k} \right) \left(1-Y_{k-1} \right) \mathbb{E} \left\lbrace U_{k-1} \vert  \overline{U}_{k-2}, \overline{O}_{k-1}, \overline{A}_{k-1}, \underline{O}_{k}, \underline{A}_{k} \right\rbrace \right]} 
\end{align*}
Applying the assumption on the missingness mechanism:
\begin{align*}
&= \frac{\mathbb{E} \left[ \frac{C_{d,k-1}}{\prod_{l=1}^{k-1}  \pi_{d,l}} \frac{ \prod_{l=1}^{k-2} U_{l}}{\prod_{l=1}^{k-1}  \pi^U_{l}} Y_{k} \left(1-Z_{k} \right) \left(1-Y_{k-1} \right) \mathbb{P} \left\lbrace U_{k-1} = 1 \vert \overline{U}_{k-2}, \overline{O}_{k-1}, \overline{A}_{k-1} \right\rbrace \right]}{\mathbb{E} \left[  \frac{C_{d,k-1}}{\prod_{l=1}^{k-1}  \pi_{d,l}} \frac{ \prod_{l=1}^{k-2} U_{l}}{\prod_{l=1}^{k-1}  \pi^U_{l}} \left(1-Z_{k} \right) \left(1-Y_{k-1} \right) \mathbb{P} \left\lbrace U_{k-1} = 1 \vert  \overline{U}_{k-2}, \overline{O}_{k-1}, \overline{A}_{k-1} \right\rbrace \right]} \\
&= \frac{\mathbb{E} \left\lbrace \frac{C_{d,k-1}}{\prod_{l=1}^{k-1}  \pi_{d,l}} \frac{ \prod_{l=1}^{k-2} U_{l}}{\prod_{l=1}^{k-2}  \pi^U_{l}} Y_{k} \left(1-Z_{k} \right) \left(1-Y_{k-1} \right) \right\rbrace}{\mathbb{E} \left\lbrace  \frac{C_{d,k-1}}{\prod_{l=1}^{k-1}  \pi_{d,l}} \frac{ \prod_{l=1}^{k-2} U_{l}}{\prod_{l=1}^{k-2}  \pi^U_{l}} \left(1-Z_{k} \right) \left(1-Y_{k-1} \right) \right\rbrace} 
\end{align*}
Applying the law of iterated expectation
\begin{align*}
&= \frac{\mathbb{E} \left[ \mathbb{E} \left\lbrace \frac{C_{d,k-1}}{\prod_{l=1}^{k-1} \pi_{d,l}} \frac{ \prod_{l=1}^{k-2} U_{l}}{\prod_{l=1}^{k-2}  \pi^U_{l}} Y_{k} \left(1-Z_{k} \right) \left(1-Y_{k-1} \right) \vert \overline{U}_{k-3}, \overline{O}_{k-2}, \overline{A}_{k-2}, \underline{O}_{k-1}, \underline{A}_{k-1} \right\rbrace \right]}{\mathbb{E} \left[ \mathbb{E} \left\lbrace \frac{C_{d,k-1}}{\prod_{l=1}^{k-1}  \pi_{d,l}} \frac{ \prod_{l=1}^{k-2} U_{l}}{\prod_{l=1}^{k-2}  \pi^U_{l}} \left(1-Z_{k} \right) \left(1-Y_{k-1} \right) \vert  \overline{U}_{k-3}, \overline{O}_{k-2}, \overline{A}_{k-2}, \underline{O}_{k-1}, \underline{A}_{k-1} \right\rbrace \right]} \\
&= \frac{\mathbb{E} \left[ \frac{C_{d,k-1}}{\prod_{l=1}^{k-1}  \pi_{d,l}} \frac{ \prod_{l=1}^{k-3} U_{l}}{\prod_{l=1}^{k-2}  \pi^U_{l}} Y_{k} \left(1-Z_{k} \right) \left(1-Y_{k-1} \right) \mathbb{E} \left\lbrace U_{k-2} \vert \overline{U}_{k-3}, \overline{O}_{k-2}, \overline{A}_{k-2}, \underline{O}_{k-1}, \underline{A}_{k-1} \right\rbrace \right]}{\mathbb{E} \left[  \frac{C_{d,k-1}}{\prod_{l=1}^{k-1}  \pi_{d,l}} \frac{ \prod_{l=1}^{k-3} U_{l}}{\prod_{l=1}^{k-2}  \pi^U_{l}} \left(1-Z_{k} \right) \left(1-Y_{k-1} \right) \mathbb{E} \left\lbrace U_{k-2} \vert  \overline{U}_{k-3}, \overline{O}_{k-2}, \overline{A}_{k-2}, \underline{O}_{k-1}, \underline{A}_{k-1} \right\rbrace \right]} 
\end{align*}
Applying the assumption on the missingness mechanism:
\begin{align*}
&= \frac{\mathbb{E} \left[ \frac{C_{d,k-1}}{\prod_{l=1}^{k-1}  \pi_{d,l}} \frac{ \prod_{l=1}^{k-3} U_{l}}{\prod_{l=1}^{k-2}  \pi^U_{l}} Y_{k} \left(1-Z_{k} \right) \left(1-Y_{k-1} \right) \mathbb{P} \left\lbrace U_{k-2} = 1 \vert \overline{U}_{k-3}, \overline{O}_{k-2}, \overline{A}_{k-2} \right\rbrace \right]}{\mathbb{E} \left[  \frac{C_{d,k-1}}{\prod_{l=1}^{k-1}  \pi_{d,l}} \frac{ \prod_{l=1}^{k-3} U_{l}}{\prod_{l=1}^{k-2}  \pi^U_{l}} \left(1-Z_{k} \right) \left(1-Y_{k-1} \right) \mathbb{P} \left\lbrace U_{k-2} = 1 \vert  \overline{U}_{k-3}, \overline{O}_{k-2}, \overline{A}_{k-2} \right\rbrace \right]} \\
&= \frac{\mathbb{E} \left\lbrace \frac{C_{d,k-1}}{\prod_{l=1}^{k-1}  \pi_{d,l}} \frac{ \prod_{l=1}^{k-3} U_{l}}{\prod_{l=1}^{k-3}  \pi^U_{l}} Y_{k} \left(1-Z_{k} \right) \left(1-Y_{k-1} \right) \right\rbrace}{\mathbb{E} \left\lbrace  \frac{C_{d,k-1}}{\prod_{l=1}^{k-1}  \pi_{d,l}} \frac{ \prod_{l=1}^{k-3} U_{l}}{\prod_{l=1}^{k-3}  \pi^U_{l}} \left(1-Z_{k} \right) \left(1-Y_{k-1} \right) \right\rbrace} 
\end{align*}
Repeating the above steps consecutively we finally obtain:
\begin{align} \label{eqnB2}
\dots = \frac{\mathbb{E} \left\lbrace \frac{C_{d,k-1}}{\prod_{l=1}^{k-1}  \pi_{d,l}} Y_{k} \left(1-Z_{k} \right) \left(1-Y_{k-1} \right) \right\rbrace}{\mathbb{E} \left\lbrace  \frac{C_{d,k-1}}{\prod_{l=1}^{k-1}  \pi_{d,l}} \left(1-Z_{k} \right) \left(1-Y_{k-1} \right) \right\rbrace} =\frac{\mathbb{E} \left\lbrace \frac{\mathbbm{1} \left( \overline{A}_{k-1} = \overline{d}_{k-1} \right)}{\prod_{l=1}^{k-1} \mathbb{P} \left(A_l = d_l \vert \overline{O}_l, \overline{A}_{l-1} = \overline{d}_{l-1} \right)} Y_{k} \left(1-Z_{k} \right) \left(1-Y_{k-1} \right) \right\rbrace}{\mathbb{E} \left\lbrace \frac{\mathbbm{1} \left( \overline{A}_{k-1} = \overline{d}_{k-1} \right)}{\prod_{l=1}^{k-1} \mathbb{P} \left(A_l = d_l \vert\overline{O}_l, \overline{A}_{l-1} = \overline{d}_{l-1} \right)} \left(1-Z_{k} \right) \left(1-Y_{k-1} \right) \right\rbrace}
\end{align}
Applying the consistency assumption
\begin{align*}
&=\frac{\mathbb{E} \left\lbrace \frac{\mathbbm{1} \left( \overline{A}_{k-1} = \overline{d}_{k-1} \right)}{\prod_{l=1}^{k-1}  \mathbb{P} \left(A_l = d_l \vert \overline{O}_l, \overline{A}_{l-1} = \overline{d}_{l-1} \right)} Y_{k}^d \left(1-Z_{k}^d \right) \left(1-Y_{k-1}^d \right) \right\rbrace}{\mathbb{E} \left\lbrace \frac{\mathbbm{1} \left( \overline{A}_{k-1} = \overline{d}_{k-1} \right)}{\prod_{l=1}^{k-1}  \mathbb{P} \left(A_l = d_l \vert\overline{O}_l, \overline{A}_{l-1} = \overline{d}_{l-1} \right)} \left(1-Z_{k}^d \right) \left(1-Y_{k-1}^d \right) \right\rbrace}
\end{align*}
Applying the law of iterated expectation
\begin{align*}
&=\frac{\mathbb{E} \left[ \mathbb{E} \left\lbrace \frac{\mathbbm{1} \left( \overline{A}_{k-1} = \overline{d}_{k-1} \right)}{\prod_{l=1}^{k-1}  \mathbb{P} \left(A_l = d_l \vert \overline{O}_l, \overline{A}_{l-1} = \overline{d}_{l-1} \right)} Y_{k}^d \left(1-Z_{k}^d \right) \left(1-Y_{k-1}^d \right) \vert  \overline{O}_{k-1}, \overline{A}_{k-2}, \underline{Y}_{k}^d, \underline{Z}_{k}^d \right\rbrace \right]}{\mathbb{E} \left[ \mathbb{E} \left\lbrace \frac{\mathbbm{1} \left( \overline{A}_{k-1} = \overline{d}_{k-1} \right)}{\prod_{l=1}^{k-1}  \mathbb{P} \left(A_l = d_l \vert \overline{O}_l, \overline{A}_{l-1} = \overline{d}_{l-1} \right)} \left(1-Z_{k}^d \right) \left(1-Y_{k-1}^d \right) \vert  \overline{O}_{k-1}, \overline{A}_{k-2}, \underline{Z}_{k}^d \right\rbrace \right]} \\
&=\frac{\mathbb{E} \left[ \frac{\mathbbm{1} \left( \overline{A}_{k-2} = \overline{d}_{k-2} \right) }{\prod_{l=1}^{k-1}  \mathbb{P} \left(A_l = d_l \vert \overline{O}_l, \overline{A}_{l-1} = \overline{d}_{l-1} \right)} Y_{k}^d \left(1-Z_{k}^d \right) \left(1-Y_{k-1}^d \right) \mathbb{E} \left\lbrace \mathbbm{1} \left( A_{k-1} = d_{k-1} \right) \vert  \overline{O}_{k-1}, \overline{A}_{k-2}, \underline{Y}_{k}^d, \underline{Z}_{k}^d \right\rbrace \right]}{\mathbb{E} \left[ \frac{\mathbbm{1} \left( \overline{A}_{k-2} = \overline{d}_{k-2} \right)}{\prod_{l=1}^{k-1}  \mathbb{P} \left(A_l = d_l \vert \overline{O}_l, \overline{A}_{l-1} = \overline{d}_{l-1} \right)} \left(1-Z_{k}^d \right) \left(1-Y_{k-1}^d \right) \mathbb{E} \left\lbrace \mathbbm{1} \left( A_{k-1} = d_{k-1} \right) \vert  \overline{O}_{k-1}, \overline{A}_{k-2}, \underline{Z}_{k}^d \right\rbrace \right]} 
\end{align*}
Applying sequential randomization
\begin{align*}
&=\frac{\mathbb{E} \left[ \frac{\mathbbm{1} \left( \overline{A}_{k-2} = \overline{d}_{k-2} \right) }{\prod_{l=1}^{k-1} \mathbb{P} \left(A_l = d_l \vert \overline{O}_l, \overline{A}_{l-1} = \overline{d}_{l-1} \right)} Y_{k}^d \left(1-Z_{k}^d \right) \left(1-Y_{k-1}^d \right) \mathbb{P} \left( A_{k-1} = d_{k-1} \vert  \overline{O}_{k-1}, \overline{A}_{k-2} = \overline{d}_{k-2} \right) \right]}{\mathbb{E} \left[ \frac{\mathbbm{1} \left( \overline{A}_{k-2} = \overline{d}_{k-2} \right)}{\prod_{l=1}^{k-1}  \mathbb{P} \left(A_l = d_l \vert \overline{O}_l, \overline{A}_{l-1} = \overline{d}_{l-1} \right)} \left(1-Z_{k}^d \right) \left(1-Y_{k-1}^d \right) \mathbb{P} \left( A_{k-1} = d_{k-1} \vert  \overline{O}_{k-1}, \overline{A}_{k-2} = \overline{d}_{k-2} \right) \right]} \\
&=\frac{\mathbb{E} \left[ \frac{\mathbbm{1} \left( \overline{A}_{k-2} = \overline{d}_{k-2} \right) }{\prod_{l=1}^{k-2}  \mathbb{P} \left(A_l = d_l \vert \overline{O}_l, \overline{A}_{l-1} = \overline{d}_{l-1} \right)} Y_{k}^d \left(1-Z_{k}^d \right) \left(1-Y_{k-1}^d \right) \right]}{\mathbb{E} \left[ \frac{\mathbbm{1} \left( \overline{A}_{k-2} = \overline{d}_{k-2} \right)}{\prod_{l=1}^{k-2}  \mathbb{P} \left(A_l = d_l \vert \overline{O}_l, \overline{A}_{l-1} = \overline{d}_{l-1} \right)} \left(1-Z_{k}^d \right) \left(1-Y_{k-1}^d \right) \right]} 
\end{align*}
Applying the law of iterated expectation
\begin{align*}
&=\frac{\mathbb{E} \left[ \mathbb{E} \left\lbrace \frac{\mathbbm{1} \left( \overline{A}_{k-2} = \overline{d}_{k-2} \right)}{\prod_{l=1}^{k-2}  \mathbb{P} \left(A_l = d_l \vert \overline{O}_l, \overline{A}_{l-1} = \overline{d}_{l-1} \right)} Y_{k}^d \left(1-Z_{k}^d \right) \left(1-Y_{k-1}^d \right) \vert  \overline{O}_{k-2}, \overline{A}_{k-3}, \underline{Y}_{k-1}^d, \underline{Z}_{k-1}^d \right\rbrace \right]}{\mathbb{E} \left[ \mathbb{E} \left\lbrace \frac{\mathbbm{1} \left( \overline{A}_{k-2} = \overline{d}_{k-2} \right)}{\prod_{l=1}^{k-2}  \mathbb{P} \left(A_l = d_l \vert \overline{O}_l, \overline{A}_{l-1} = \overline{d}_{l-1} \right)} \left(1-Z_{k}^d \right) \left(1-Y_{k-1}^d \right) \vert  \overline{O}_{k-2}, \overline{A}_{k-3}, \underline{Y}_{k-1}^d, \underline{Z}_{k-1}^d\right\rbrace \right]} \\
&=\frac{\mathbb{E} \left[ \frac{\mathbbm{1} \left( \overline{A}_{k-3} = \overline{d}_{k-3} \right) }{\prod_{l=1}^{k-2}  \mathbb{P} \left(A_l = d_l \vert \overline{O}_l, \overline{A}_{l-1} = \overline{d}_{l-1} \right)} Y_{k}^d \left(1-Z_{k}^d \right) \left(1-Y_{k-1}^d \right) \mathbb{E} \left\lbrace \mathbbm{1} \left( A_{k-2} = d_{k-2} \right) \vert  \overline{O}_{k-2}, \overline{A}_{k-3}, \underline{Y}_{k-1}^d, \underline{Z}_{k-1}^d \right\rbrace \right]}{\mathbb{E} \left[ \frac{\mathbbm{1} \left( \overline{A}_{k-3} = \overline{d}_{k-3} \right)}{\prod_{l=1}^{k-2}  \mathbb{P} \left(A_l = d_l \vert \overline{O}_l, \overline{A}_{l-1} = \overline{d}_{l-1} \right)} \left(1-Z_{k}^d \right) \left(1-Y_{k-1}^d \right) \mathbb{E} \left\lbrace \mathbbm{1} \left( A_{k-2} = d_{k-2} \right) \vert  \overline{O}_{k-2}, \overline{A}_{k-3}, \underline{Y}_{k-1}^d, \underline{Z}_{k-1}^d \right\rbrace \right]} 
\end{align*}
Applying sequential randomization
\begin{align*}
&=\frac{\mathbb{E} \left[ \frac{\mathbbm{1} \left( \overline{A}_{k-3} = \overline{d}_{k-3} \right) }{\prod_{l=1}^{k-2} \mathbb{P} \left(A_l = d_l \vert \overline{O}_l, \overline{A}_{l-1} = \overline{d}_{l-1} \right)} Y_{k}^d \left(1-Z_{k}^d \right) \left(1-Y_{k-1}^d \right) \mathbb{P} \left( A_{k-2} = d_{k-2} \vert  \overline{O}_{k-2}, \overline{A}_{k-3} = \overline{d}_{k-3} \right) \right]}{\mathbb{E} \left[ \frac{\mathbbm{1} \left( \overline{A}_{k-3} = \overline{d}_{k-3} \right)}{\prod_{l=1}^{k-2}  \mathbb{P} \left(A_l = d_l \vert \overline{O}_l, \overline{A}_{l-1} = \overline{d}_{l-1} \right)} \left(1-Z_{k}^d \right) \left(1-Y_{k-1}^d \right) \mathbb{P} \left( A_{k-2} = d_{k-2} \vert  \overline{O}_{k-2}, \overline{A}_{k-3} = \overline{d}_{k-3} \right) \right]} \\
&= \frac{\mathbb{E} \left[ \frac{\mathbbm{1} \left( \overline{A}_{k-3} = \overline{d}_{k-3} \right) }{\prod_{l=1}^{k-3}  \mathbb{P} \left(A_l = d_l \vert \overline{O}_l, \overline{A}_{l-1} = \overline{d}_{l-1} \right)} Y_{k}^d \left(1-Z_{k}^d \right) \left(1-Y_{k-1}^d \right) \right]}{\mathbb{E} \left[ \frac{\mathbbm{1} \left( \overline{A}_{k-3} = \overline{d}_{k-3} \right)}{\prod_{l=1}^{k-3}  \mathbb{P} \left(A_l = d_l \vert \overline{O}_l, \overline{A}_{l-1} = \overline{d}_{l-1} \right)} \left(1-Z_{k}^d \right) \left(1-Y_{k-1}^d \right) \right]} \\
& = \dots =\frac{\mathbb{E} \left[ Y_{k}^d \left(1-Z_{k}^d \right) \left(1-Y_{k-1}^d \right) \right]}{\mathbb{E} \left[ \left(1-Z_{k}^d \right) \left(1-Y_{k-1}^d \right) \right]} =\frac{\mathbb{E} \left[ Y_{k}^d \mathbbm{1} \left(Z_{k}^d = 0\right) \mathbbm{1} \left(Y_{k-1}^d = 0\right) \right]}{\mathbb{E} \left[ \mathbbm{1} \left(Z_{k}^d = 0\right) \mathbbm{1} \left(Y_{k-1}^d = 0\right) \right]} \\
&= \mathbb{P} \left( Y_{k}^d = 1 \vert  Z_{k}^d = 0, Y_{k-1}^d = 0 \right) = h_k^{(1)}(d).
\end{align*}
We obtain the claim of the lemma from the fact that under no propensity score model misspecification 
\begin{equation*}
\frac{\sum_{i=1}^n Y_{k,i} \left( 1 - Z_{k,i} \right) \left( 1 - Y_{k-1,i} \right) \hat{W}_{d,k-1,i}}{\sum_{i=1}^n  \left( 1 - Z_{k,i} \right) \left( 1 - Y_{k-1,i} \right) \hat{W}_{d,k-1,i}}
\end{equation*}
consistently estimates
\begin{equation*}
\frac{\mathbb{E} \left\lbrace Y_{k} \left( 1 - Z_{k} \right) \left( 1 - Y_{k-1} \right) W_{d,k-1} \right\rbrace}{\mathbb{E} \left\lbrace \left( 1 - Z_{k} \right) \left( 1 - Y_{k-1} \right) W_{d,k-1} \right\rbrace}.
\end{equation*}
\\
Similarly, for $h_k^{(2)}(d)$. 
\end{proof}

\begin{lemma}
\label{lemma3}
We define the following estimating functions
\begin{align*}
U^{(1)} \left( \alpha, \bf{q^{(1)} } \right) & \coloneqq \sum_d \sum_{k=1}^{K} \left\lbrace Y_{k} - h_{k}^{(1)} \left(d, V; \alpha \right) \right\rbrace \left( 1 - Y_{k-1} \right) \left( 1 - Z_{k} \right) q_k^{(1)}  \left(d, V \right) W_{d,k-1} \\
U^{(2)} \left( \beta, \bf{q^{(2)} } \right) & \coloneqq \sum_d \sum_{k=1}^{K} \left\lbrace Z_{k} - h_{k}^{(2)} \left(d, V; \beta \right) \right\rbrace \left( 1 - Z_{k-1} \right) \left( 1 - Y_{k-1} \right) q_k^{(2)}  \left(d, V \right) W_{d,k-1}
\end{align*}
where $q_k^{(1)}  \left(d, V \right)$ is an arbitrary $\mathbb{R}^{\dim(\alpha)}$-valued function and $q_k^{(2)}  \left(d, V \right)$ is an arbitrary $\mathbb{R}^{\dim(\beta)}$-valued function. For an arbitrary $\mathbb{R}^{\dim(\alpha) \times K}$-valued function $\mathbf{q^{(1)}}$  and an arbitrary $\mathbb{R}^{\dim(\beta) \times K}$-valued function $\mathbf{q^{(2)}}$ 
\begin{align*}
\mathbb{E} \left\lbrace U^{(1)}  \left( \alpha, \mathbf{q^{(1)} } \right) \right\rbrace &= 0 \\
\mathbb{E} \left\lbrace U^{(2)}  \left( \beta, \mathbf{q^{(2)} } \right) \right\rbrace &= 0.
\end{align*}
\end{lemma}

\begin{proof}[Proof of Lemma~\ref{lemma3}]
\begin{equation*}
\mathbb{E} \left\lbrace U_1 \left( \alpha, \bf{q^{(1)}} \right) \right\rbrace = \mathbb{E} \left[ \sum_d \sum_{k=1}^{K} \left\lbrace Y_{k} - h_k^{(1)} \left(d, V; \alpha \right) \right\rbrace \left( 1 - Y_{k-1} \right) \left( 1 - Z_{k} \right) q_k^{(1)} \left(d, V \right) W_{d,k-1} \right] 
\end{equation*}
Under correct specification of the marginal structural model
\begin{align*}
&= \mathbb{E} \left[ \sum_d \sum_{k=1}^{K} \left\lbrace Y_{k} - h_{k}^{(1)} \left(d, V \right) \right\rbrace \left( 1 - Y_{k-1} \right) \left( 1 - Z_{k} \right) q_k^{(1)} \left(d, V \right) W_{d,k-1} \right] \\
&= \sum_d \sum_{k=1}^{K} \mathbb{E} \left[ \left\lbrace Y_{k} - h_k^{(1)} \left(d, V \right) \right\rbrace \left( 1 - Y_{k-1} \right) \left( 1 - Z_{k} \right) q_k^{(1)} \left(d, V \right) W_{d,k-1} \right]
\end{align*}
Applying law of iterated expectation
\begin{align*}
&= \sum_d \sum_{k=1}^{K} \mathbb{E} \left( \mathbb{E} \left[ \left\lbrace Y_{k} - h_k^{(1)} \left(d, V \right) \right\rbrace \left( 1 - Y_{k-1} \right) \left( 1 - Z_{k} \right) q_k^{(1)} \left(d, V \right) W_{d,k-1} \vert  V \right] \right) \\
&= \sum_d \sum_{k=1}^{K} \mathbb{E} \left( \mathbb{E} \left[ \left\lbrace Y_{k} - h_k^{(1)} \left(d, V \right) \right\rbrace \left( 1 - Y_{k-1} \right) \left( 1 - Z_{k} \right) W_{d,k-1} \vert  V \right] q_k^{(1)} \left(d, V \right) \right) \\
&= \sum_d \sum_{k=1}^{K} \mathbb{E} \left( \left[ \mathbb{E} \left\lbrace Y_{k} \left( 1 - Y_{k-1} \right) \left( 1 - Z_{k} \right) W_{d,k-1} \vert  V \right\rbrace - h_k^{(1)} \left(d, V \right) \mathbb{E} \left\lbrace \left( 1 - Y_{k-1} \right) \left( 1 - Z_{k} \right) W_{d,k-1} \vert  V \right\rbrace \right] \right. \\
&\left. \times q_k^{(1)} \left(d, V \right) \right) \\
&= \sum_d \sum_{k=1}^{K} \mathbb{E} \left( \left[ \mathbb{E} \left\lbrace Y_{k} \left( 1 - Y_{k-1} \right) \left( 1 - Z_{k} \right) W_{d,k-1} \vert  V \right\rbrace  - \frac{\mathbb{E} \left\lbrace Y_{k} \left( 1 - Y_{k-1} \right) \left( 1 - Z_{k} \right) W_{d,k-1} \vert  V \right\rbrace}{\mathbb{E} \left\lbrace \left( 1 - Y_{k-1} \right) \left( 1 - Z_{k} \right) W_{d,k-1} \vert  V \right\rbrace} \right. \right.\\
&\left. \left.\times \mathbb{E} \left\lbrace \left( 1 - Y_{k-1} \right) \left( 1 - Z_{k} \right) W_{d,k-1} \vert  V \right\rbrace \right] q_k^{(1)} \left(d, V \right) \right) \\
&= 0. 
\end{align*}
Similarly, for $U_2 \left( \beta, \bf{q^{(2)}} \right)$. 
\end{proof}

\end{document}